\documentstyle[graphics,preprint,aps]{revtex}
\begin{document}
\title{Magnetic Ordering and Superconductivity in\\ 
the RE$_2$Ir$_3$Ge$_5$ (RE$=$ Y, La-Tm, Lu) System}
\author{Yogesh Singh and S. Ramakrishnan}
\address{Tata Institute Of Fundamental Research,Bombay-400005, India}
\maketitle
\begin{abstract}
We report structure, electrical resistivity, magnetic susceptibility, isothermal 
magnetization and heat-capacity studies on polycrystalline samples of the intermetallic series
RE$_2$Ir$_3$Ge$_5$ (RE~$=$ Y, La, Ce-Nd, Gd-Tm, Lu) from 1.5 to 300~K. We find that the compounds for RE~$=$ Y, La-Dy,  
crystallize in the tetragonal Ibam (U$_2$Co$_3$Si$_5$ type) structure whereas the compounds 
for RE$=$ Er-Lu, crystallize in a new orthorhombic structure with a space group Pmmn. Samples of Ho$_2$Ir$_3$Ge$_5$ were always found to be multiphase.
The compounds for RE~=~Y to Dy which adopt the Ibam type structure show a metallic resistivity whereas the compounds with RE~=~Er, Tm and Lu show an anomalous behavior in the resistivity with a semiconducting increase in $\rho$ as we go down in temperature from 300~K. Interestingly we had earlier found a positive temperature coefficient of resistivity for the Yb sample in the same temperature range. We will compare this behavior with similar observations in the compounds RE$_3$Ru$_4$Ge$_{13}$ and REBiPt.
La$_2$Ir$_3$Ge$_5$ and Y$_2$Ir$_3$Ge$_5$ show bulk superconductivity below 1.8~K and 2.5~K respectively. 
Our results confirm that Ce$_2$Ir$_3$Ge$_5$ shows a Kondo lattice behavior and undergoes 
antiferromagnetic ordering below 8.5~K. 
Most of the other compounds containing magnetic rare-earth elements undergo a single
antiferromagnetic transition at low temperatures (T$\leq$12~K) while Gd$_2$Ir$_3$Ge$_5$, Dy$_2$Ir$_3$Ge$_5$ and Nd$_2$Ir$_3$Ge$_5$ show multiple transitions. 
The T$_N$'s for most of the compounds roughly scale with the de Gennes factor.
which suggests that the primary mechanism of interaction leading to the magnetic ordering of the  magnetic moments may be the RKKY interaction. The ordering temperature of 8.5~K for Ce$_2$Ir$_3$Ge$_5$ is anomalously large compared to the T$_N$ for Gd$_2$Ir$_3$Ge$_5$ which is about 12~K.There are signs of strong CEF influence on the measured properties for the series.\\

\normalsize
\noindent
Ms number ~~~~~~~~~~~~~~~~~~~~PACS number:74.70.Ad, 74.25.Bt, 74.25.Ha
\end{abstract}

\newpage
\section{Introduction}
\label{sec:INTRO}
\noindent
Rare earth ternary silicides and germanides of the type RE$_2$T$_3$X$_5$, where T is a transition metal and X is either Si or Ge have been extensively investigated for their unusual magnetic and superconducting properties and the rich variety of structures they form in \cite{r1,r2,r3,r4,r5}.
In particular, compounds belonging to the RE$_2$Fe$_3$Si$_5$ series have prompted considerable efforts to understand their superconductivity and magnetism \cite{r5,r6,r7,r8}. Both Tm$_2$Fe$_3$Si$_5$ and Lu$_2$Fe$_3$Si$_5$ compounds show superconductivity at low temperatures. In the Tm sample, superconductivity at about 1.5~K is destroyed by the onset of anti-ferromagnetic order at 1~K and it re-enters the normal state \cite{r9} whereas Lu$_2$Fe$_3$Si$_5$ has the highest T$_C$= 6~K for an iron containing compound \cite{r10}. Recently it was reported that Er$_2$Fe$_3$Si$_5$ was also found to be superconducting below 1~K \cite{r11}.
In this structure, Fe does not have any moment and it only helps in building up the large density of states at the Fermi level. Recently we have established Yb$_2$Fe$_3$Si$_5$ to be a heavy fermion compound with Kondo-Lattice behavior and anti-ferromagnetic ordering below 1.7~K \cite{r12}. Thus it is clear that compounds of the series RE$_2$Fe$_3$Si$_5$ exhibit unusual superconducting and magnetic properties.
The RE$_2$Ir$_3$Ge$_5$ series of compounds form in a structure closely related to the RE$_2$Fe$_3$Si$_5$ structure. The compound Ce$_2$Ir$_3$Ge$_5$ of this series has been studied in some detail by various groups \cite{r13,r14,r15} but there have been little efforts to make a detailed study of the other compounds of the series RE$_2$Ir$_3$Ge$_5$.
Recently we had succeded in preparing and studying the compound Yb$_2$Ir$_3$Ge$_5$. We found that apart from showing interesting low temperature properties \cite{r16} it also formed in a structure different from it's Ce analogue. This prompted us to make a comprehensive study of the structural and magnetic properties of the complete RE$_2$Ir$_3$Ge$_5$ series to look for systematic trends and variations in the physical properties across the series.
Here we report our detailed resistivity, magnetic susceptibility, isothermal magnetization and heat-capacity results for the series RE$_2$Ir$_3$Ge$_5$ (RE$=$Y, La, Ce-Nd, Gd-Tm, Lu) from 1.5 to 300~K. 
\section{EXPERIMENTAL DETAILS} 
\label{sec:EXPT}
\noindent 
Samples of RE$_2$Ir$_3$Ge$_5$  (RE$=$Y, La-Nd, Gd-Tm,Lu) were made by melting the 
individual constituents (taken in stoichiometric proportions) 
in an arc furnace under Ti gettered high-purity argon atmosphere on a water cooled copper hearth. The purity of the rare-earth metals and Ir was 99.9\% whereas the purity of Ge was 99.999\%. 
The alloy buttons were flipped over and remelted five to six times to ensure homogenous mixing. 
The samples were annealed at 950~$^0$C  for a period of 10 days before slowly cooling down to room temperature. The X-ray powder diffraction pattern of the samples did not show the presence of any parasitic impurity phases. The samples with RE$=$ Y, La, Ce-Dy, were found to adopt the tetragonal Ibam (U$_2$Co$_3$Si$_5$ type) structure as reported earlier for the compound Ce$_2$Ir$_3$Ge$_5$ \cite{r15} whereas the compounds 
for RE$=$ Er-Lu, crystallize in an orthorhombic structure with a space group Pmmn which we had recently reported \cite{r16} for the sample Yb$_2$Ir$_3$Ge$_5$. The X-ray pattern for the sample Ho$_2$Ir$_3$Ge$_5$ always showed many extra reflections which shows that Ho was the transition point for the structural change. Although the X-ray for the sample Er$_2$Ir$_3$Ge$_5$ did not show any extra peaks, preliminary microanalysis study showed the presence of a small amount of second phase. 
The lattice constants a, b and c are given in Table 1 where they are seen to 
decrease across the series (i.e., from Ce$_2$Ir$_3$Ge$_5$ to 
Dy$_2$Ir$_3$Ge$_5$). 
The U$_2$Co$_3$Si$_5$ structure has 
only a single site for U whereas the Pmmn structure allows two sites for the rare-earth element.  
Hence, the net hybridization between d- and f- orbitals can be very 
different for these two structures which has to be taken into account in the 
analysis of magnetic ordering temperatures. 
\\
The temperature dependence  of dc susceptibility ($\chi$) was measured using
a commercial SQUID magnetometer in the temperature range from 1.8 to 300~K. 
The ac susceptibility was measured using a home-built susceptometer
\cite{r17} from 1.5 to 20~K. The absolute accuracy with which magnetization 
measurements were performed is within 1\%. The resistivity was measured using
an ac resistance bridge (Linear Research Inc., USA) from 1.5 to 300~K.
We used a four-probe dc technique with contacts  made using silver paint on thin slides cut from the annealed samples. The temperature was 
measured using a calibrated Si diode (Lake~Shore~Inc., USA) sensor. 
All the data were collected using an IBM compatible 
PC/AT via IEEE-488 interface. The relative accuracy of the resistance
measurements is 50~ppm while the accuracy of the absolute resistivity is
only 5\% due to errors in estimating the geometrical factors. The 
heat-capacity in zero field between 1.7 and  30~K was measured with an accuracy
of 1\% using an automated adiabatic heat pulse calorimeter. A calibrated 
germanium resistance thermometer (Lake~Shore~Inc, USA) was used as the 
temperature sensor in this range. 
\section{RESULTS}
\label{sec:RD} 
\subsection{Magnetic susceptibility  and Magnetization studies}
\subsubsection{Properties of Y$_2$Ir$_3$Ge$_5$, La$_2$Ir$_3$Ge$_5$ and Lu$_2$Ir$_3$Ge$_5$} 
\label{sec:SUS1}
\noindent 
The temperature dependence of the dc magnetic susceptibility ($\chi$)
of La$_2$Ir$_3$Ge$_5$, Y$_2$Ir$_3$Ge$_5$ and  Lu$_2$Ir$_3$Ge$_5$ are shown
in Fig.~1. The inset for La$_2$Ir$_3$Ge$_5$ clearly shows an abrupt diamagnetic drop just below 1.75~K which marks the onset of the sample into the superconducting state. In the normal state the $\chi$ of La$_2$Ir$_3$Ge$_5$ is diamagnetic between 75 to 300~K but shows a Curie-Wiess like tail as we go down in temperature. It is possible that there is a small amount of magnetic impurity in La or Ir metals. However, the susceptibility $\chi$ for both the Y and Lu samples are practically temperature independent down to low temperatures with only small upturns at the lowest temperatures (probably due to paramagnetic impurities) hence the possibility of Ir having a significant amount of impurity is remote. For Y$_2$Ir$_3$Ge$_5$, a somewhat sharper diamagnetic transition below 2.5~K as seen in the low temperature inset for this compound signals the transition into the superconducting state for this compound. 
We did not observe any diamagnetic signal down to 1.8~K for Lu$_2$Ir$_3$Ge$_5$.

\subsubsection{Properties of RE$_2$Ir$_3$Ge$_5$ (RE$=$Ce-Dy)}
\label{sec:SUS2}
\noindent 
The temperature dependence of the inverse dc magnetic susceptibility
($\chi$$^{-1}$) of RE$_2$Ir$_3$Ge$_5$ (RE$=$Ce, Pr, and Nd)
is shown in Fig.~2. Similar data for RE$=$Gd,Tb and Dy are shown in Fig.~3. Insets show the low temperature $\chi$ behavior of the respective compounds. The high
temperature susceptibility (100~K $<$ T $<$ 300~K) could be fitted to a
modified Curie-Weiss expression which is given by,
$$ \chi~=~\chi_0~+~{C \over (T-\theta_p)}~, \eqno(1) $$
where C is the Curie constant which can be written in terms of the effective
moment as,
$$ C~(emu~K/mol)~\approx~{\mu_{eff}(\mu_B)^2~x \over 8}~, \eqno(2) $$
where $x$ is the number of magnetic RE ions per formula unit.
The values of $\chi_0$, $\mu_{eff}$, and $\theta_p$ are given in Table 2. The 
main contributions to the temperature independent $\chi_0$ are namely, the 
diamagnetic susceptibility which arises due to the presence of ion cores and 
the Pauli spin susceptibility of the conduction electrons. The estimated 
effective moment in all the cases is found to be quite close to the free ion moment 
of RE$^{3+}$ ion telling us that we are dealing with trivalent rare-earth ions here. For most of the compounds (except Ce$_2$Ir$_3$Ge$_5$), we get a relatively small and negative value of the Curie-Weiss temperature ($\theta_p$). A negetive $\theta_p$ implies the presence of antiferromagnetic 
correlations. For Ce$_2$Ir$_3$Ge$_5$, we obtain a negative and large value of $\theta_p$~=~-137~K indicating a strong hybridization of 4f orbitals of Ce. This is in agreement with an earlier report \cite{r15} which also estimated a large value of -160~K for $\theta_p$. 
The value of $\theta_p$ for most of the other samples is between -10 to -20~K which although smaller than the value for Ce$_2$Ir$_3$Ge$_5$, is still somewhat larger than e.g. the values obtained for the compounds of the series RE$_2$Rh$_3$Sn$_5$ \cite{r18}.
The insets clearly show the antiferromagnetic ordering for all the compounds. For Ce$_2$Ir$_3$Ge$_5$, we also observe an abrupt upturn in the susceptibility below 4.5~K after the sample has undergone an anti-ferromagnetic transition at 8.5~K. This upturn in $\chi$ below T$_N$ was also observed in an earlier work \cite{r19}. There are no signatures of this transition in the resistivity but there is a weak kink in the heat capacity data. This feature seen in both the magnetic susceptibility and heat capacity measurements could be intrinsic to the sample or may be due to a smal ferromagnetic impurity phase. Microanalysis studies are needed to verify whether this behavior is intrinsic to the sample.
The low temperature $\chi$ data for Gd$_2$Ir$_3$Ge$_5$ also shows a second shoulder like feature around 4.5~K apart from the anti-ferromagnetic transition at about 12~K. 
The transition temperatures of various compounds obtained from magnetic susceptibility measurements are listed in Table 3. The ordering temperatures have been determined from peaks/inflection points in the d($\chi$T)/dT vs T plots. The d($\chi$T)/dT vs T plots for Nd$_2$Ir$_3$Ge$_5$ and Dy$_2$Ir$_3$Ge$_5$ showed additional anomalies apart from the single transition visible in the $\chi(T)$ data. This can be seen in Fig.~4 where we show the d($\chi$T)/dT vs T plots for Nd$_2$Ir$_3$Ge$_5$ and Dy$_2$Ir$_3$Ge$_5$. Thus it can be seen that Nd$_2$Ir$_3$Ge$_5$ shows two anomalies at 2.1 and 2.82~K while Dy$_2$Ir$_3$Ge$_5$ shows three anomalies at 2, 4.3 and 7.2~K. We will later show that these multiple transitions are also observed in the resistivity and heat capacity measurements. 

\subsubsection{Properties of RE$_2$Ir$_3$Ge$_5$ (RE$=$Er and Tm)}
\label{sec:SUS3}
\noindent 
The plots of the inverse magnetic susceptibility as a function of temperature for the magnetic compounds forming in the different Pmmn structure ie RE$_2$Ir$_3$Ge$_5$ (
RE$=$Er and Tm) is shown in fig.~5. The insets show the low temperature susceptibility $\chi$ data. Curie-Wiess fits to the high temperature (100 $<$ T $<$ 300~K) inverse susceptibility ($\chi$$^{-1}$) data gives values of the effective moments which are close to the RE$^{3+}$ values. The parameters obtained from the fit are given in table 2 along with the values for the other compounds.
We did not observe any signature of magnetic ordering for both Tm$_2$Ir$_3$Ge$_5$ and Er$_2$Ir$_3$Ge$_5$ down to 1.8~K although the d($\chi$T)/dT vs T plot for the Er$_2$Ir$_3$Ge$_5$ sample shown as the second inset in the top panel of Fig.~5 shows a minimum just around 2~K which could be a possible signature of magnetic order. This is corroborated by anomalies at the same temperature in the resistivity and heat capacity data which we will discuss below. 

\subsubsection{Magnetization studies of  RE$_2$Ir$_3$Ge$_5$ (RE$=$Ce, Nd, Gd and Dy 
)}
\label{sec:Mag1}
\noindent 
Isothermal magnetization measurements at temperatures both above and below the Neel temperature T$_N$ have been performed on some of the samples of the series RE$_2$Ir$_3$Ge$_5$. Fig.~6 shows the magnetization curves for the samples Ce$_2$Ir$_3$Ge$_5$, Nd$_2$Ir$_3$Ge$_5$, Gd$_2$Ir$_3$Ge$_5$ and Dy$_2$Ir$_3$Ge$_5$.
We observe a non-linear behavior in M vs H at 2~K ($<$ T$_N$) for all the samples. This non-linear behavior  agrees with the notion of antiferromagnetic ordering of RE$^{3+}$ spins. 
This non-linear behavior persists upto 7~K in the case of Ce compound. At higher temperatures (T $>$ T$_N$), one observes the usual linear behavior in magnetization which characterizes the paramagnetic state. 
The magnetization values of Ce are very small presumably due to the presence 
of Kondo effect. The magnetization data for Gd$_2$Ir$_3$Ge$_5$ at 2~K starts out linearly upto 10~KOe but shows a slight upturn at higher fields which continues upto 60~KOe with no sign of saturation. At 20~K the magnetization is linear upto the highest fields with a slightly smaller slope than the curve at 2~K.
The magnetization data of Nd and Dy at 2~K show an S type of shape with increasing field indicating possible metamagnetic transitions. The magnetization for Nd$_2$Ir$_3$Ge$_5$ at 2~K starts out linearly but begins to show an upturn starting at 15~KOe which continues upto a field of 40~KOe after which it shows signs of saturating at higher fields reaching a value of 2.77$\mu_B$ which is slightly lower than the free ion value of 3.62$\mu_B$ for Nd$^3{+}$. Higher fields may be required to reach the full moment value since the magnetization has not completely saturated even for the highest field (~=~60~KOe) used in our measurements. This metamagnetic upturn is also seen in the magnetization curve at 3~K. At 15~K, we get the linear behavior expected in the paramagnetic state. The magnetization for Dy$_2$Ir$_3$Ge$_5$ at 2~K and 5~K starts out with a slightly sublinear behaviour at low fields upto about 10~KOe after which it begins to curve upwards upto a field of 25~KOe. For higher fields it sort of saturates, reaching a value of about 10.4$\mu$$_B$ at 60~KOe which is close to the free moment value for Dy$^{3+}$ ions. A similar behavior for the magnetization with saturation at 50~KOe is seen at a temperature of 15~K which is much above the Neel temperature of 7~K for Dy$_2$Ir$_3$Ge$_5$. A linear behavior for the magnetization is seen only at 35~K. Clearly there are short range magnetic correlations even above T$_N$ in this compound and this aspect would be worth further investigation prefarably by neutron scattering experiments. 

\subsection{Resistivity studies on RE$_2$Ir$_3$Ge$_5$ (RE$=$Y, La-Tm)}
\label{sec: res1}
The resistivity data ($\rho$) for La$_2$Ir$_3$Ge$_5$ and Y$_2$Ir$_3$Ge$_5$ is shown in fig.~7. The insets show the low temperature data which clearly show a sharp drop in the resistivity marking the onset of the superconducting transitions for both the samples below 1.7~K and 2.4~K respectively.
 This drop corresponds with the diamagnetic signal observed in the susceptibility measurement for both the samples. An earlier report \cite{r15} did not find any superconductivity for La$_2$Ir$_3$Ge$_5$. However, their investigations had been carried out only down to 1.8~K below which we observe the resistive drop and diamagnetic signal in the susceptibility.
The resistivity ($\rho$) data for the RE$_2$Ir$_3$Ge$_5$ (RE~=~ Ce-Dy) series is shown in fig.~8 and fig.~9. The insets for all the samples shows the low temperature data on an expanded scale.
From the insets one can see that the resistivity data of all the magnetic rare earth samples except that of Pr$_2$Ir$_3$Ge$_5$  
show a change of slope at their respective magnetic transitions. For the Dy$_2$Ir$_3$Ge$_5$ sample, one can see from the inset in fig.~9 that there is a second feature which appears as a shoulder at about 4.5~K below the anti-ferromagnetic transition at 8~K. There is a similar but much weaker feature in the $\rho$ data for Gd$_2$Ir$_3$Ge$_5$ below the main transition at 12~K. This corresponds to the second peak observed in the susceptibility data for this sample. However, there is only one distinct change of slope in the low temperature $\rho$ data for Nd$_2$Ir$_3$Ge$_5$ whereas we had seen two anomalies in the d($\chi$)/dT vs T plots and which are also seen in the heat capacity data discussed later. Plotting d($\rho$)/dT vs T brings out the third transition in Dy$_2$Ir$_3$Ge$_5$ and the second transition in Gd$_2$Ir$_3$Ge$_5$ more disticntly as shown in fig.~10. However, taking  d($\rho$)/dT for Nd$_2$Ir$_3$Ge$_5$ did not help because it was very noisy and possibly the two transitions are very close together in temperature and may not be individually distinguishable in the resistivity data.
Fig.~11 shows the temperature dependence of $\rho$ for Er$_2$Ir$_3$Ge$_5$, Tm$_2$Ir$_3$Ge$_5$ and Lu$_2$Ir$_3$Ge$_5$ which crystallize in the orthogonal Pmmn structure. It is evident that the $\rho$ behavior for these compounds is anomalous showing a semiconducting rise of $\rho$ for all three compounds as we cool down from 300~K. It is interesting to note that we had found a positive temperature coeffecient of the resistivity for the sample Yb$_2$Ir$_3$Ge$_5$ \cite{r16} in the same temperature range. Similar behavior for the resistivity has been observed earlier for the series of compounds R$_3$Ru$_4$Ge$_{13}$ \cite{r20} and RBiPt \cite{r21} where the resistivity for all samples except Yb showed a semiconducting response.
The Er$_2$Ir$_3$Ge$_5$ data is highly anomalous. It initially increases with decreasing temperature like it's isostructural neihbours Tm and Lu but then it reaches a broad maximum at nearly 100~K before starting to decrease more or less linearly with temperature. The resistivity then shows an upturn at about 5~K similar to that seen in Kondo lattice systems before decreasing abruptly below 2~K indicating a possible magnetic ordering. Evidence of magnetic order has earlier been seen in the d$\chi$/dT vs T plot which showed an anomaly at 2~K and we will later see that the heat capacity for this sample also shows a peak around 2~K. 
There is no evidence of any ordering in the resistivity data for Tm$_2$Ir$_3$Ge$_5$ down to 1.8~K. There is an abrupt drop of almost 70\% in the $\rho$ for Lu$_2$Ir$_3$Ge$_5$ below 2~K which may possibly be the onset of superconductivity although the resistivity does not fall completely to zero. Our heat capacity data also shows an incomplete anomaly just below 2~K as will be discussed later. The transition temperatures observed from resistivity data are compared with those obtained from susceptibility and heat-capacity studies (described later) in Table 3.\\   
In the normal state \textit{i.e.} above the superconducting/magnetic transition temperatures, we have been able to fit the low temperature (T$_C$ or T$_N$ to 25~K) dependence of $\rho$ to a power law which can be written as,
$$  \rho~=~\rho_0~+~a~T^n \eqno(3) $$
The values of $\rho_0$, $a$, and n are given in Table~4.
For La$_2$Ir$_3$Ge$_5$ and Y$_2$Ir$_3$Ge$_5$, the optimum value of n was
found to be 3.6 and 3.3, respectively. These values do not agree with Wilson's 
s-d scattering model which predicts a T$^3$ dependence of $\rho$(T) for 
T$<$~$\theta_D$/10. The discrepancy could arise due to a variety of reasons such as,
complex structure of the Fermi surface, phonon-drag effects and lattice 
anharmonicity.\\ 
The $\rho$ of most of the magnetic rare-earth samples (except Ce$_2$Ir$_3$Ge$_5$) shows a power law 
dependence in the low temperature paramagnetic region (T$_N$$<$T$<$25~K) with n different from 2 in all cases except for Gd$_2$Ir$_3$Ge$_5$. The value of n for both Tb$_2$Ir$_3$Ge$_5$ and Dy$_2$Ir$_3$Ge$_5$ is nearly equal to 2.1 however, the values for Pr$_2$Ir$_3$Ge$_5$ (n~=~2.4) and Nd$_2$Ir$_3$Ge$_5$ (n~=~1.7) deviate markedly from the T$^2$ dependence and are not understood at present.\\ 
At high temperatures (75~K$<$T$<$300~K), the resistivity data  deviate 
significantly from the expected linear temperature dependence. This
has been seen in many compounds where the value of $\rho$ becomes sufficiently
large for the mean free path to shorten to the order of a few atomic spacings.
When that happens, the scattering cross section will no longer
be linear in the scattering perturbation. Since the dominant temperature-
dependent scattering mechanism is electron-phonon interaction here, the
$\rho$ will no longer be proportional to the mean square atomic displacement,
which is proportional to T for a harmonic potential. Instead, the
resistance will rise less rapidly than linearly in T and will show negative
curvature (d$^2$$\rho$/dT$^2$~$<$0). This behavior was  also seen in our previous
studies on silicides and germanides \cite{r4,r18}. 

One of the models which describe the high temperature $\rho$(T) of these compounds is known 
as the parallel resistor model \cite{r22}. In this model the expression of 
$\rho$(T) is given by, 
$$ {1\over{\rho(T)}} =
{1\over{\rho_1(T)}} + {1\over{\rho_{max}}} ~, \eqno(4) $$ 
where $\rho_{max}$ is the temperature independent saturation resistivity and $\rho_1$(T) is the ideal temperature-dependent
resistivity. Further, the ideal resistivity is given by the following
expression,  $$ \rho_1(T) =
\rho_0 +C_1({T\over{\theta_D}})^3\int^{\theta_D/T}_0
{x^3dx\over{(1-exp(-x))(exp(x)-1)}} ~, \eqno(5) $$ where $\rho_0$ is the
residual resistivity and the second term is due to phonon-assisted
electron scattering similar to the s-d scattering in transition metal
compounds. $\theta_D$ is the Debye temperature and $C_1$ is a numerical
constant. 
The high temperature (75~K$<$T$<$300~K) $\rho$ data of the samples with RE~=~Y, La-Dy (which show metallic behavior)
could be fitted nicely to the parallel resistor model. The values of the various 
parameters obtained from the fit to this model are listed in Table~5.
The observed $\theta_D$ values from the heat-capacity data (discussed
below) are also given in Table 5. No attempt was made to fit the parallel
resistor model to the $\rho$ data of Ce$_2$Ir$_3$Ge$_5$ since this
compound exhibits Kondo behavior.

\subsection{Heat capacity studies on RE$_2$Ir$_3$Ge$_5$ (RE$=$La-Tm,Lu,Y)}
\label{sec:hc1}
The temperature dependence of the heat-capacity ($C_p$) from 1.8 to 30~K of
various compounds of the series RE$_2$Ir$_3$Ge$_5$ are shown in Fig.~12, Fig.~13 and
Fig. 14. Also shown in the same figures are the magnetic contribution ($C_{mag}$)to the heat capacity and the estimated entropy ($S_{mag}$).The insets show the $C_p/T$ vs $T^2$ data at
low Temperatures.
For the magnetic samples RE$_2$Ir$_3$Ge$_5$ (RE~=~Ce-Dy), we had a choice to use either La$_2$Ir$_3$Ge$_5$ or Y$_2$Ir$_3$Ge$_5$ as the nonmagnetic counterpart to estimate the lattice contribution. We have used La$_2$Ir$_3$Ge$_5$ for the purpose since we find that in some cases the lattice contribution is overestimated if we use Y$_2$Ir$_3$Ge$_5$. For the samples with RE~=~ Er and Tm, we have used Lu$_2$Ir$_3$Ge$_5$ as the reference sample. 
The temperature dependence of C$_p$ of the La, Y, Lu and some of the magnetic samples above their magnetic transition temperatures could be 
fitted to the expression,
$$ C_p~=~\gamma~T~+~\beta~T^3~, \eqno(7) $$ where $\gamma$ is due
to the electronic contribution and $\beta$ is due to the lattice contribution.
From the $\beta$ value, we can estimate the value of the Debye temperature $\theta_D$ using the relation,
$$ \theta_D~=~({12~\pi^4~N~r~k_B \over 5 \beta})^{1/3}~, \eqno(8) $$
where N is the Avogadro's number, r is the number of atoms per formula
unit, and k$_B$ is the Boltzmann's constant. 
We find that for the La and Y compounds the estimated $\gamma$ values are 16~mJ/RE~mol~K$^2$ and 19.4~mJ/RE~mol~K$^2$ respectively which are quite high for non-magnetic samples and could possibly indicate a large density of states at the Fermi level. The absence of superconductivity in Lu$_2$Ir$_3$Ge$_5$ above 2~K is attributed to its relatively low $\gamma$ value which we found to be 9~mJ/Lu~mol~K$^2$.

The temperature dependence of C$_p$ from 1.8 to 30~K of Ce$_2$Ir$_3$Ge$_5$
is shown in Fig.~12. The inset shows the C$_p$/T vs T$^2$ data at low 
temperatures. The large peak seen at 8.3~K in the inset confirms the bulk ordering of the Ce$^{3+}$ moments. This temperature is comparable to the values of the transition temperature as obtained by the $\chi$ and $\rho$(T) measurements. 
This transition temperature also closely resembles the previously reported values \cite{r15}.
It must be noted that we observe a small shoulder at 4.3~K in the C$_P$ data where we had observed a strong ferromagnetic upturn in the $\chi$ data.  
The extrapolated value for the Sommerfeld's electronic heat-capacity coefficient $\gamma$ was found to be 
188 mJ/Ce~mol~K$^2$ which classifies it as a moderately heavy fermion system. However, estimation of $\gamma$ from data above T$_N$ can be easily influenced by magnetic correlations and CEF effects and may not be strictly correct. For a correct estimation of the true value of $\gamma$, data down to much lower temperatures would be required.
The estimated entropy at 30~K is found to be 8.83~J/~Ce~mol~K which is much less than the expected value of 
Rln(2J+1) (with J~=~5/2 for Ce). The reduced value of the entropy implies that there are higher lying CEF levels which have not been populated at these temperatures and so the whole entropy is not released. The Kondo effect seen in the resistivity data could also be partly responsible for the reduced entropy.
The experimentally obtained values of the entropy and the expected values have been compared for all compounds in Table 6.

The temperature dependence of C$_p$ from 1.8 to 30~K of Pr$_2$Ir$_3$Ge$_5$
is also shown in Fig.~12. The inset clearly shows a distinct peak at a T$^2$ value corresponding to a temperature just above 2~K. This anomaly in the heat capacity, along with the peak in the susceptibility data at a similar temperature clearly establishes bulk antiferromagnetic ordering for the compound although we did not see any change in slope in the resistivity data down to 1.7~K.
We observe a broad hump around 7~K in the C$_{P}$ data. This is more prominent in the magnetic heat capacity C$_{mag}$ obtained by subtracting the lattice part. This could be a Schottky type anomaly indicating the presence of low lying excited CEF levels which become populated as the temperature is increased. Similar behaviour has been observed in many other Pr based compounds like Pr$_2$Rh$_3$Si$_5$ \cite{r4} and Pr$_2$Rh$_3$Sn$_5$ \cite{r18}. For the former case the authors have used a singlet ground state and a doublet excited state for the CEF levels to explain the data. 
The estimated entropy of 15.8~J/Pr~mol~K at 30~K is again found to be less than the expected
value of Rln(2J+1) clearly suggesting the influence of CEF levels at these temperatures.

The temperature dependence of C$_p$ from 1.8 to 30~K of Nd$_2$Ir$_3$Ge$_5$ (Fig. 12) shows two distinct and seperate anomalies (seen more clearly in the C$_P$/T vs T$^2$ inset) at 2~K and 2.8~K corroborating the two anomalies seen earlier in the d$\chi$/dT plot. The estimated entropy in this case is also much less than the expected value of Rln(2J+1) (see Table 6) which implies the presence of CEF contributions. 

The C$_p$ for Gd$_2$Ir$_3$Ge$_5$ from 1.8 to 30~K is shown in Fig.~13. We observe a large $\lambda$ type anomaly at 11.2~K which clearly indicates bulk magnetic ordering of Gd$^{3+}$ moments. A broad shoulder is also visible at 4.2~K which corresponds with the second peak seen earlier in the $\chi$ and d$\rho$/dT data at roughly the same temperature. This second anomaly can be associated with the way the (2J+1) multiplet under consideration evolves within the ordered state. This low temperature hump following a magnetic transition at a higher temperature has been seen in some other Gd based compounds such as GdBiPt \cite{r21} and GdCu$_2$Si$_2$) \cite{r23}.    
The inset with the low temperature C$_P$/T vs T$^2$ data shows the two anomalies more clearly.
The estimated entropy at 30~K is found to be 16.8~J/Gd~mol~K which is 
nearly equal to the expected value of Rln(2J+1). Note that at T$_N$ the entropy has already reached 88$\%$ of its value at 30~K.

The C$_p$ data for Tb$_2$Ir$_3$Ge$_5$ also shown in Fig.~13 shows a huge (10~J/Tb~mol) peak at the magnetic ordering temperature of 6~K. The magnetic heat capacity shows what looks like the low temperature tail of a Schottky like hump around 29~K. The reduced entropy value at 30~K is indicative of CEF effects being important at these temperatures.

The C$_P$ vs T data for Dy$_2$Ir$_3$Ge$_5$ shown in the same figure shows three distinct anomalies at 2.1~K, 4.8~K and 7.4~K. We had also seen three anomalies at roughly these temperatures in the d$\chi$/dT and d$\rho$/dT plots for this sample. These features below the first main transition could be due to reorientation of the spins in the ordered state. Usually the change is small enough to escape a distinct detection in a magnetic measurement. However, in the reorientation of spins, some degree of freedom is involved and hence a signature in the heat capacity. Reorientation of spins is just a conjecture at the present time and it is possible that this compound actually has a complicated magnetic structure at low temperatures. This issue can be settled with neutron diffraction to probe the low temperature magnetic structure and the changes it undergoes across the three transitions.
The last panel in the same figure shows the C$_P$ data for Y$_2$Ir$_3$Ge$_5$. The inset showing the C$_P$/T vs T$^2$ data at low temperatures shows an anomaly peaked at T$^2$~=~6~K$^2$ which corresponds to the superconducting transition seen at 2.5~K in the $\rho$ and $\chi$ measurements. 
The value of $\delta$C/$\gamma$T$_C$ is found to be .64 which is much reduced from the value 1.43 for a BCS type superconductor. This indicates that Y$_2$Ir$_3$Ge$_5$ is a very weakely coupled superconductor.
 
Fig.~14 shows the heat capacity data for the compounds RE$_2$Ir$_3$Ge$_5$ (RE~=~Er, Tm, Lu) forming in the structure different from the rest of the compounds.
The C$_P$ data for Er$_2$Ir$_3$Ge$_5$ shows an upturn at low temperatures starting at 6~K and undergoes a maximum peaked around 1.9~K. However, we could not trace the complete transition down to lower temperatures because of experimental limitations. This corresponds with the anomalies seen in the d$\chi$/dT and $\rho$ data discussed earlier.

The C$_P$ data for Tm$_2$Ir$_3$Ge$_5$ shown in the same figure also shows an upturn below 3.5~K which continues down to 1.8~K. This may be the onset of the magnetic ordering of Tm$^{3+}$ moments in this sample which we have not been able to capture because of the transition being below 2~K. The estimated entropy of only 9.2 J/Tm~mol~K at 30~K is much reduced from the expected value of ln(2J+1) (with J~=~6 for Tm). However, it must be noted that we see a strong indication that the compound may order below 2~K and a lot of entropy would be sitting under the peak at the transition when it occurs.
The heat capacity data for Lu$_2$Ir$_3$Ge$_5$ is also shown in Fig.~14. The C$_P$ vs T$^2$ inset shows an incomplete anomaly around 2~K which corresponds with the abrupt drop in resistivity at the same temperature and could be a signature of superconductivity in this compound although we could not observe any diamagnetic signal in our magnetic measurements down to 1.8~K.

\section{Discussion}
\label{sec:DIS}
In this section we will make an attempt to understand the temperature
dependence of the measured physical properties and the models which we
have used to understand their behavior and look for systematic trends followed across the series. We begin with the susceptibility behaviour. The high temperature data for all samples could be fitted to a modified Curie-Wiess law. The extracted effective moments for all samples are close to their theoretical values for free RE$^{3+}$ ions showing that we are dealing with trivalent moments here and that there is no contribution from the Ir. It must be recalled that the Yb$_2$Ir$_3$Ge$_5$ sample showed a much reduced moment estimated from the high temperature data and that was attributed to the fact that the rare-earth element has two inequivalent sites in the crystal structure and so the Yb could have a different valence at the two sites \cite{r16}. We see here that both the Er and Tm samples, which form in the same structure as the Yb sample show trivalent behaviour with Er$_2$Ir$_3$Ge$_5$ undergoing magnetic ordering below 2~K as seen in the heat capacity and resistivity measurements while Tm$_2$Ir$_3$Ge$_5$ is also seen to be on the verge of magnetic order and we can already see the onset in the low temperature heat capacity inset for the sample. From table~2 it can be seen that the temperature independent susceptibility $\chi_0$ is non-negligible for some cases which possibly indicates a large density of states at the Fermi level N(E$_F$).  
The values of $\theta_P$ are found to be of the order of -15~K (see table 2) for most of the compounds. These values are significantly higher than those of RE$_2$Rh$_3$Sn$_5$ \cite{r18} for example. This explains the higher values of the T$_N$'s of the compounds of the RE$_2$Ir$_3$Ge$_5$ series.
The data begins to deviate from the Curie-Wiess behaviour at lower temperatures because of the influence of crystalline electric feilds and because the magnetic correlations begin to grow.
In the section where we described the results from susceptibility measurements we mentioned that we have used the d($\chi$T)/dT vs T plots to determine the antiferromagnetic transition temperatures. Near an antiferromagnetic transition the temperature dependence of d($\chi$T)/dT mimicks the magnetic heat capacity curve \cite{r24}. This was demonstrated beautifully in Fig~4 where we had plotted d($\chi$T)/dT vs T for Nd$_2$Ir$_3$Ge$_5$ and Dy$_2$Ir$_3$Ge$_5$. Comparing this curve with the magnetic heat capacity of Nd$_2$Ir$_3$Ge$_5$ and Dy$_2$Ir$_3$Ge$_5$, one can see a clear similarity in the shape of the curves near the various transitions. The transition temperatures however, can be deduced unambigously by this method only by taking into account data from other measurements also because taking the derivative sometimes gives some spurious peaks and in general the noise is enhanced in the derivative and so one has to be careful in determining T$_N$ by this method.  
We now turn our attentions to the resistivity data. From table~4 it can be seen that most of the samples have resistivity values typical of rare-earth intermetallic compounds at low temperatures.
The power law fit to the low temperature data for La$_2$Ir$_3$Ge$_5$ and Y$_2$Ir$_3$Ge$_5$ in their normal state as described in the section III B shows that the data deviates from the expected T$^3$ dependence predicted by the Wilsons s-d scattering model. There are cases of many non-magnetic intermetallic alloys where the low temperature resistivity data deviates from the T$^3$ dependence and follows a power law behaviour with n$<$3. However, it is difficult to find many compounds showing n$>$3 as we find for both La$_2$Ir$_3$Ge$_5$ and Y$_2$Ir$_3$Ge$_5$. The reasons are not well understood at present but could be due to lattice anharmonicity or phonon drag effects.
Attempts to fit the resistivity data in the paramagnetic region for the compounds containing magnetic rare-earth show that only Gd$_2$Ir$_3$Ge$_5$ follows a T$^2$ dependence suggesting dominance of scattering by spin fluctuations at these temperatures for this compound. Both Tb$_2$Ir$_3$Ge$_5$ and Dy$_2$Ir$_3$Ge$_5$ show a power law behaviour with n~=~2.1 which is close to the T$^2$ dependence. However, for both Pr$_2$Ir$_3$Ge$_5$ and Nd$_2$Ir$_3$Ge$_5$ we find a marked deviation from the T$^2$ law with n being equal to 2.4 and 1.7 respectively. These T$^{2.4}$ and T$^{1.7}$ dependences for $\rho$ at low temperatures is quite puzzling and not understood at present.
The Ce$_2$Ir$_3$Ge$_5$ compound shows heavy fermion behaviour and the expected T$^2$ behaviour might occur at very low temperatures. 
We could also fit the data below T$_N$ for many magnetic samples to a power law dependence. The values of n$>$1 found for most of the samples except Gd$_2$Ir$_3$Ge$_5$ (n$<$1) is expected below the anti-ferromagnetic ordering. The fractional dependence of $\rho$ below T$_N$ for Gd$_2$Ir$_3$Ge$_5$ can be attributed to the scattering of conduction electrons by critical spin fluctuations which begin to grow as one approaches T$_N$. Such a behaviour has also been observed earlier in other Gd based samples such as Gd$_2$Rh$_3$Sn$_5$ \cite{r18}.
One clearly needs more studies on these compounds to try to understand their transport properties and clean single crystals would be helpful in doing so because it is well known that in ternary silicides and germanides the transport properties can be highly anisotropic and the overall behaviour for a polycrystalline sample can be easily influenced by this.   
The semiconducting behavior in the resistivity for the samples Er$_2$Ir$_3$Ge$_5$, Tm$_2$Ir$_3$Ge$_5$ and Lu$_2$Ir$_3$Ge$_5$ is interesting. In our recent report on the sample Yb$_2$Ir$_3$Ge$_5$ we had found a metallic resistivity in the same temperature range \cite{r16}. A similar behavior in the transport properties has been observed in R$_3$Ru$_4$Ge$_{13}$ \cite{r20} and RBiPt \cite{r21} where the resistivity for all samples except Yb showed a semiconducting response. However, one major difference between the behavior we observe is that there is no evidence for a gap or pseudogap as seen in these compounds since our data does not follow an activated behavior for $\rho$(T). This is shown in fig.~15 where we have plotted ln($\rho$) vs 1/T for the Er, Tm and Lu sample in the region where we observe a semiconducting behavior. In this respect the $\rho$(T) behavior for our samples is similar to that of URh$_2$Ge$_2$ \cite{r25} where a negative temperature coefficient of resistivity is found upto room temperature for some samples. We believe that like in the case of URh$_2$Ge$_2$, the anomalous resistivity behavior is arising due to crystallographic disorder which occurs due to inter-site exchange between Ir and Ge. Further investigations are definitely required to understand this behavior.  
We could fit the high temperature dependence of $\rho$ to
the parallel resistor model (see Table~5) successfully and the $\theta_D$
values obtained from such fits agree roughly with those obtained from
heat-capacity data for almost all compounds (magnetic or non-magnetic). 
The reasons for the difference of about 10 to 20\% could be due to
anharmonic contribution to the transport which is not considered in the parallel
resistor model. The values of $\rho_{max}$ also vary considerably across
the series and are quite high in some cases. Thus it is clear that both the low temperature and the high temperature behaviour of the transport properties for these compounds require more investigations for a better understanding.

The parameters obtained from the analysis of the heat capacity measurements can be found in Table~6 where we have listed the values of the ordering temperature T$_N$, entropy S$_{mag}$(T$_N$)/R, J,
ln(2J+1) and S(30~K)/R. From the column giving values of S$_{mag}$(T$_N$)/R one can see that for Ce$_2$Ir$_3$Ge$_5$, Tb$_2$Ir$_3$Ge$_5$ and Er$_2$Ir$_3$Ge$_5$, the entropy just above the transition reaches a value which is close to $ln(2)$ indicating that the ground state for these compounds is a doublet. The entropy for Ce$_2$Ir$_3$Ge$_5$ increases only weakely after the transition upto about 30~K indicating that the ground state doublet is well seperated from the excited crystal field levels.
The entropy values at T$_N$ for Nd$_2$Ir$_3$Ge$_5$ and Dy$_2$Ir$_3$Ge$_5$ reach values of 1.01~R (~$\approx$~Rln3) and 1.4~R (~$\approx$~Rln4) respectively which indicates that the ground state is a triplet and a quartet for these systems respectively. The entropy at T$_N$ for Gd$_2$Ir$_3$Ge$_5$ is already 1.8~R and reaches almost the full 2.08~R at 30~K. The entropy at T$_N$ for Pr$_2$Ir$_3$Ge$_5$  
is unusually low at .05~R and is quite puzzling. The entropy reaches almost R~ln(3) at 10~K after the Schottky anomaly which is peaked at 5~K. This would have been consistent with a nonmagnetic CEF groundstate with a doublet forming the first excited state or maybe two singlets close together. However, we do see a magnetic transition in both the susceptibility and heat capacity measurements. The reason for this unusually small entropy at the transition is not understood at present.
The entropy for most of the samples continues to rise above T$_N$, indicating the participation of excited crystal field levels in this temperature range, but reach values considerably reduced from the full R~ln(2~J+1).

In general, if CEF effects are neglected,
the antiferromagnetic ordering temperature T$_M$  for a series of 
isostructural and isoelectronic metals are expected to scale 
(de Gennes scaling \cite{r26}) as (g$_J$$-$1)$^2$~J(J+1) where g$_J$ is the Lande g factor and J is the total angular momentum of the local moment. If the angular momentum is quenched
then T$_N$s are expected to scale as S(S+1).

The solid line in Fig.~16 is the dG factor (g$_J$~-~1)$^2$~J(J+1) normalized to the value for Gd. The dashed line is obtained by similar normalization
for the case where the orbital angular momentum L is quenched and S is the good quantum number. From the Fig.~16, it is
evident that the ordering temperatures (higest transition temperature for samples with multiple transitions) of the compounds  roughly follow
the de Gennes scaling (g$_J$-1)$^2$~J(J+1)). The slight difference is probably due to CEF effects which are quite strong as we saw in our heat capacity measurements. The fact that most of the compounds follow the de Gennes scaling implies that the main interaction leading to the magnetic transitions in this series may be the RKKY interaction. It is worthwhile to note that the T$_N$ for Ce (~=8.5~K) is anomalously large compared to the other compounds. 
\section{CONCLUSION}
\label{sec:CON}
To conclude, we have synthesized and studied compounds of the series RE$_2$Ir$_3$Ge$_5$ with RE~=~Y, La, Ce-Nd, Gd-Tm and Lu using X-ray powder diffraction, magnetic susceptibity, isothermal magnetization, electrical resistivity and heat capacity measurements. We find that the structure changes from a tetragonal U$_2$Co$_3$Si$_5$ type structure for Y, La and Ce-Dy to a new orthorhombic structure with space group Pmmn for Er-Lu. The non-magnetic compounds La$_2$Ir$_3$Ge$_5$ and Y$_2$Ir$_3$Ge$_5$ show superconductivity below 1.8~K and 2.4~K respectively while for the compound Lu$_2$Ir$_3$Ge$_5$ indications for superconductivity could be seen in the resistivity and heat capacity measurements only. The absence of bulk superconductivity above 2~K for this compound may be attributed to the low density of states at the fermi level for the latter compound as indicated by the small value of the Sommerfield's coefficient for this compound compared to the Y and La compounds.
All compounds containing magnetic rare-earth elements were found to give an estimated effective moment $\mu$$_{eff}$ close to the free ion RE$^{3+}$ values and show magnetic ordering below 12~K or onset of magnetic order as in the case of Tm$_2$Ir$_3$Ge$_5$. Dy$_2$Ir$_3$Ge$_5$, Nd$_2$Ir$_3$Ge$_5$ and Gd$_2$Ir$_3$Ge$_5$ show multiple transitions apart from the main anti-ferromagnetic transition. The ordering temperature of Ce$_2$Ir$_3$Ge$_5$ at 8.5~K is anomalously high compared to the other compounds considering that Gd$_2$Ir$_3$Ge$_5$ orders at 12~K. Ce$_2$Ir$_3$Ge$_5$ shows a Kondo-lattice behavior with a doublet ground state and moderate heavy electron behaviour. 
The T$^{3.6}$ and T$^{3.3}$ power law behaviour of the normal state $\rho$ data for La$_2$Ir$_3$Ge$_5$ and Y$_2$Ir$_3$Ge$_5$ respectively and the T$^{2.4}$ and T$^{1.7}$ power law dependence of the $\rho$ data in the paramagnetic state for Pr$_2$Ir$_3$Ge$_5$ and Nd$_2$Ir$_3$Ge$_5$ is not understood at present. We find a semiconducting resistivity for the compounds Er$_2$Ir$_3$Ge$_5$, Tm$_2$Ir$_3$Ge$_5$ and Lu$_2$Ir$_3$Ge$_5$ which we believe is arising due to crystallographic disorder caused by an inter-site exchange between the Ir and Ge.
The transport properties for this series of compounds clearly merits and requires further investigations on cleaner samples and preferably on single crystals to investigate the role of anisotropy on the overall behaviour of $\rho$. 
From the temperature dependence of the entropy for the various compounds we have been able to establish that the ground state for the compounds Ce$_2$Ir$_3$Ge$_5$, Tb$_2$Ir$_3$Ge$_5$ and Er$_2$Ir$_3$Ge$_5$ is a doublet while the ground state for Nd$_2$Ir$_3$Ge$_5$ and Dy$_2$Ir$_3$Ge$_5$ is a triplet and a quartet respectively. We could also observe the complete octuplet for Gd$_2$Ir$_3$Ge$_5$. It was difficult to establish the ground states for the Pr$_2$Ir$_3$Ge$_5$ and the Tm$_2$Ir$_3$Ge$_5$ compounds given that the latter compound does not order down to the lowest temperatures of our measurements and the former compound shows an anomalously low entropy above the magnetic transition.
Finally, the transition temperatures for most of the compounds scale with the de Genne's factor indicating that the chief mechanism through which the magnetic moments interact may actually be the RKKY type.

\newpage

\begin{figure}
\begin{center}\resizebox{4.5in}{!}{\includegraphics{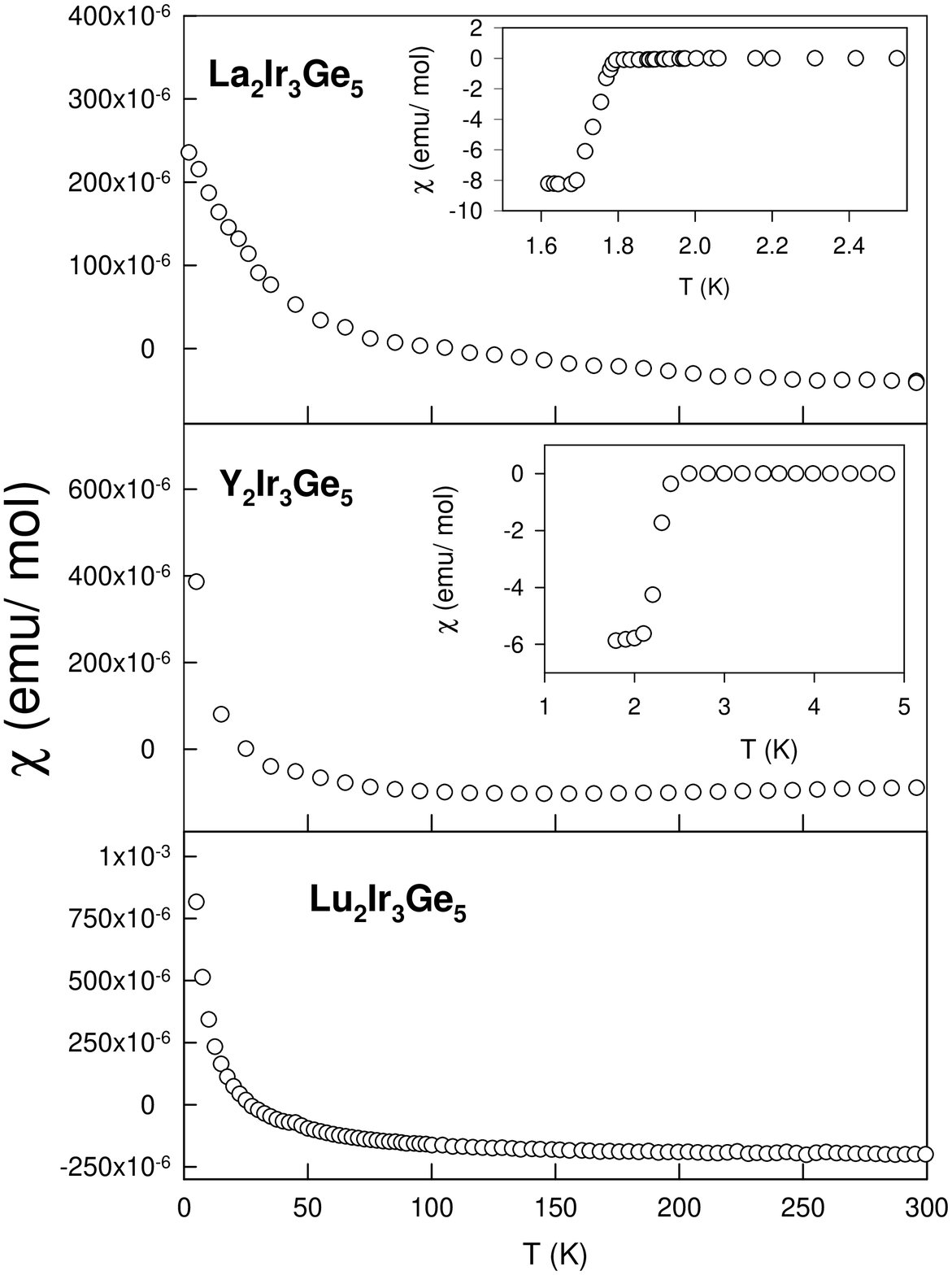}}\end{center}
\caption{Variation of susceptibility ($\chi$) of La$_2$Ir$_3$Ge$_5$, Y$_2$Ir$_3$Ge$_5$ and Lu$_2$Ir$_3$Ge$_5$
from 1.8 to 300~K in a field of 1~kOe. The insets for La$_2$Ir$_3$Ge$_5$ and Y$_2$Ir$_3$Ge$_5$ show the $\chi$ data down to 1.5~K in a field of 10~Oe showing the diamagnetic drop at the respective superconducting transition temperatures. No diamagnetism was seen for Lu$_2$Ir$_3$Ge$_5$. \label{fsus1}}
\end{figure} 
\begin{figure} 
\begin{center}\resizebox{4.5in}{!}{\includegraphics{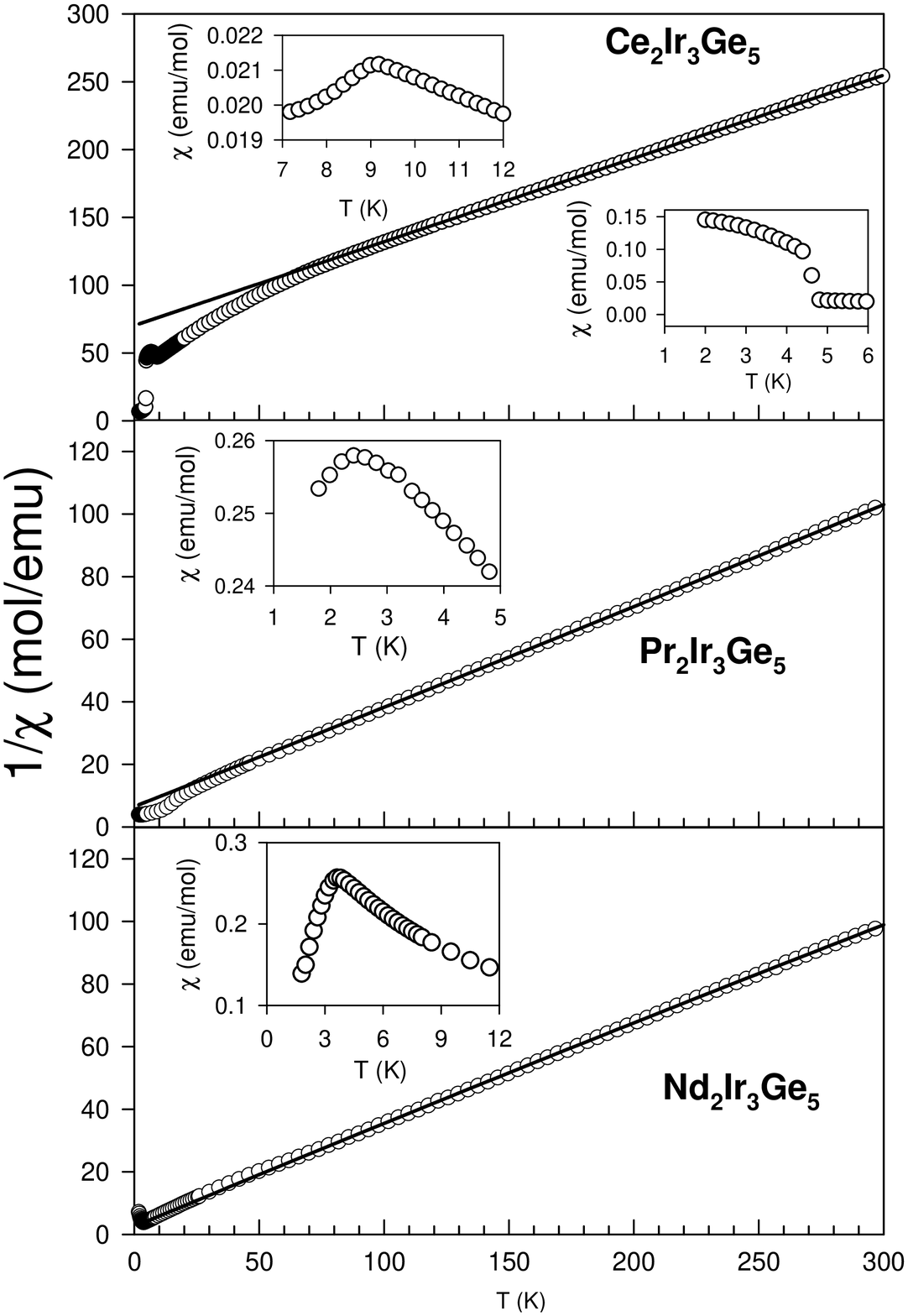}}\end{center}
\caption{Variation of inverse dc susceptibility (1/$\chi$) of 
RE$_2$Ir$_3$Ge$_5$ (RE$=$Ce, Pr and Nd) from 1.8 to 300~K in a field of 1~kOe. 
The inset shows the $\chi$ vs T behavior at low temperatures.
The solid line is a fit to the Curie-Weiss relation 
(see text for details). \label{fsus2}}
\end{figure} 
\begin{figure} 
\begin{center}\resizebox{4.5in}{!}{\includegraphics{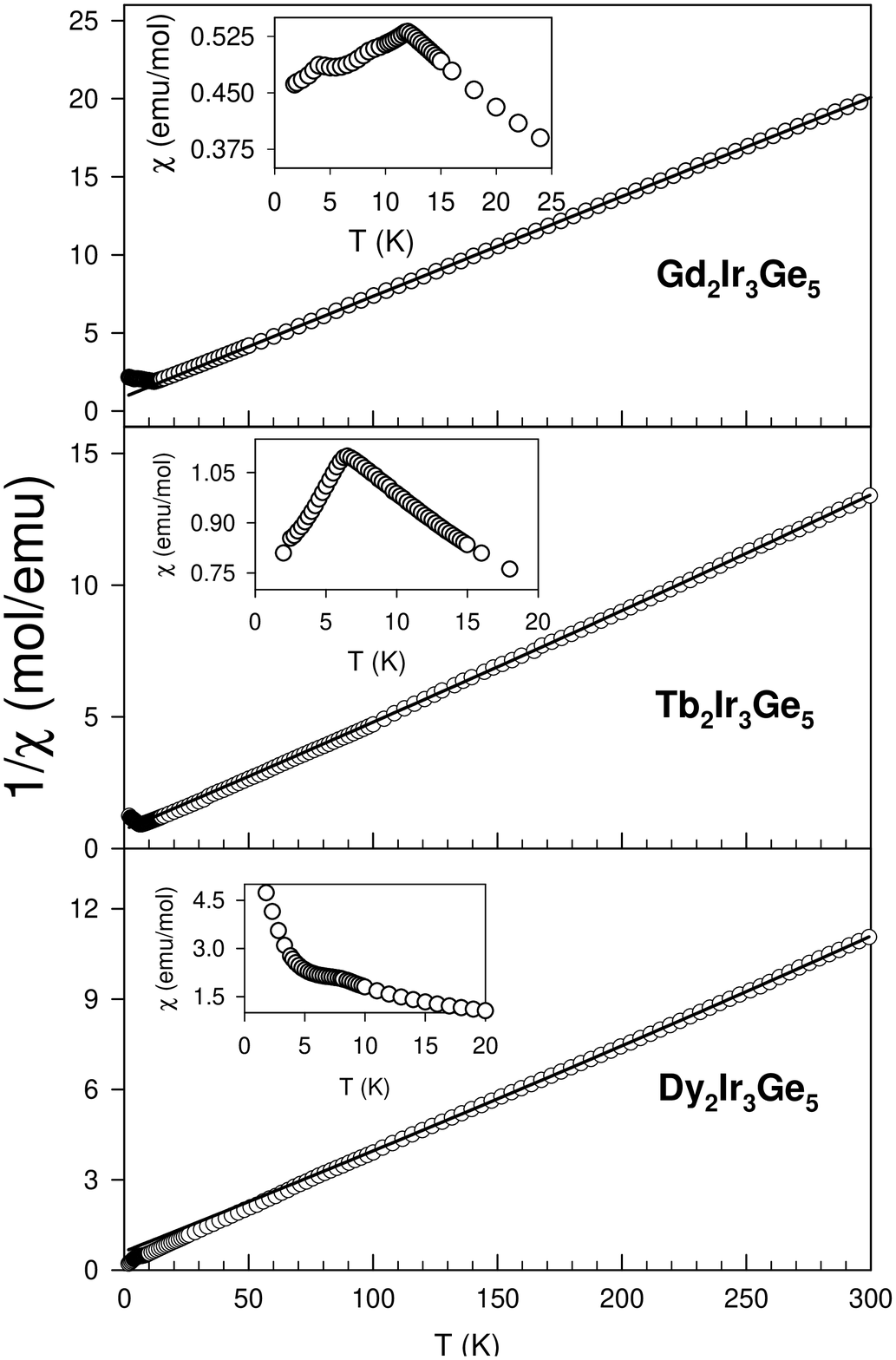}}\end{center}
\caption{Variation of inverse dc susceptibility (1/$\chi$) of 
RE$_2$Ir$_3$Ge$_5$ (RE$=$Gd, Tb and Dy)  from 1.8 to 300~K in a field of 
1~kOe. The inset shows the $\chi$ vs T behavior at low temperatures.
The solid line is a fit to the Curie-Weiss relation 
(see text for details). \label{fsus3}}
\end{figure}
\begin{figure} 
\begin{center}\resizebox{4.5in}{!}{\includegraphics{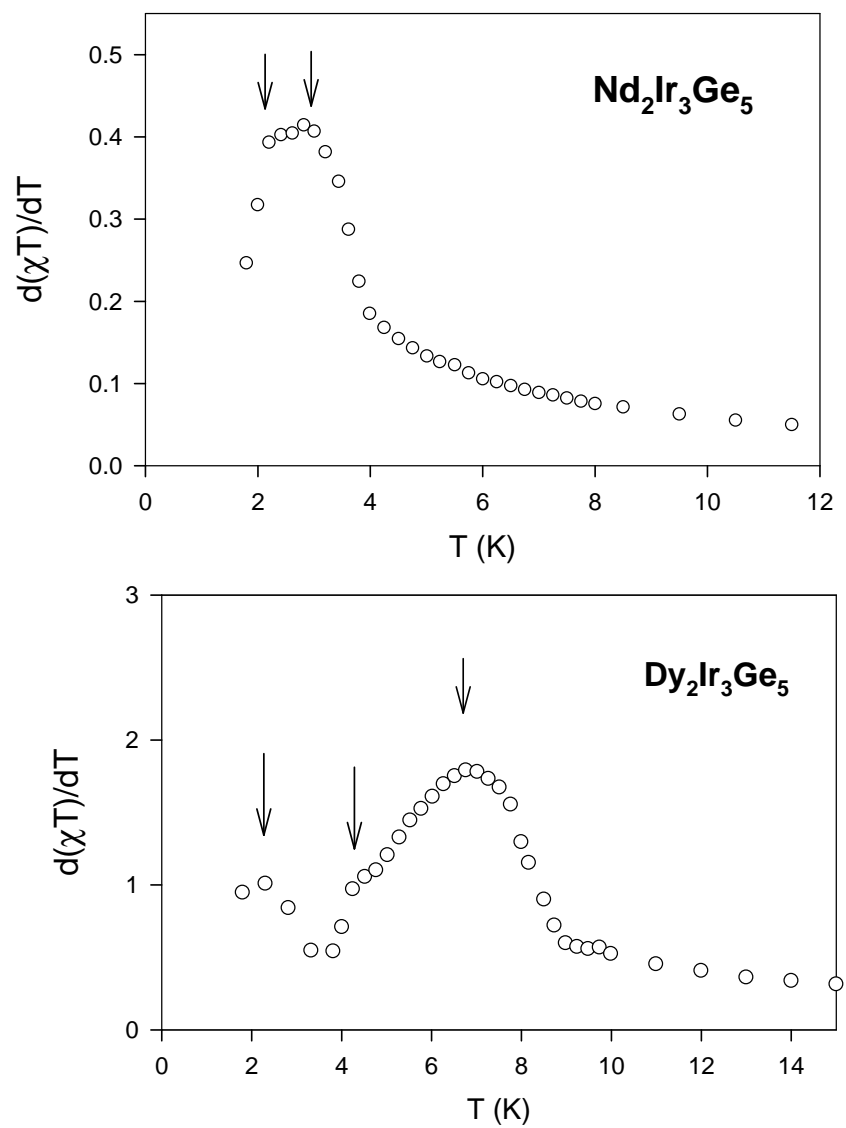}}\end{center}
\caption{d$\chi$T/dT vs T for Nd$_2$Ir$_3$Ge$_5$ and Dy$_2$Ir$_3$Ge$_5$. The arrows shows the multiple anomalies for both compounds.} \label{fsus4}
\end{figure} 
\begin{figure} 
\begin{center}\resizebox{4.5in}{!}{\includegraphics{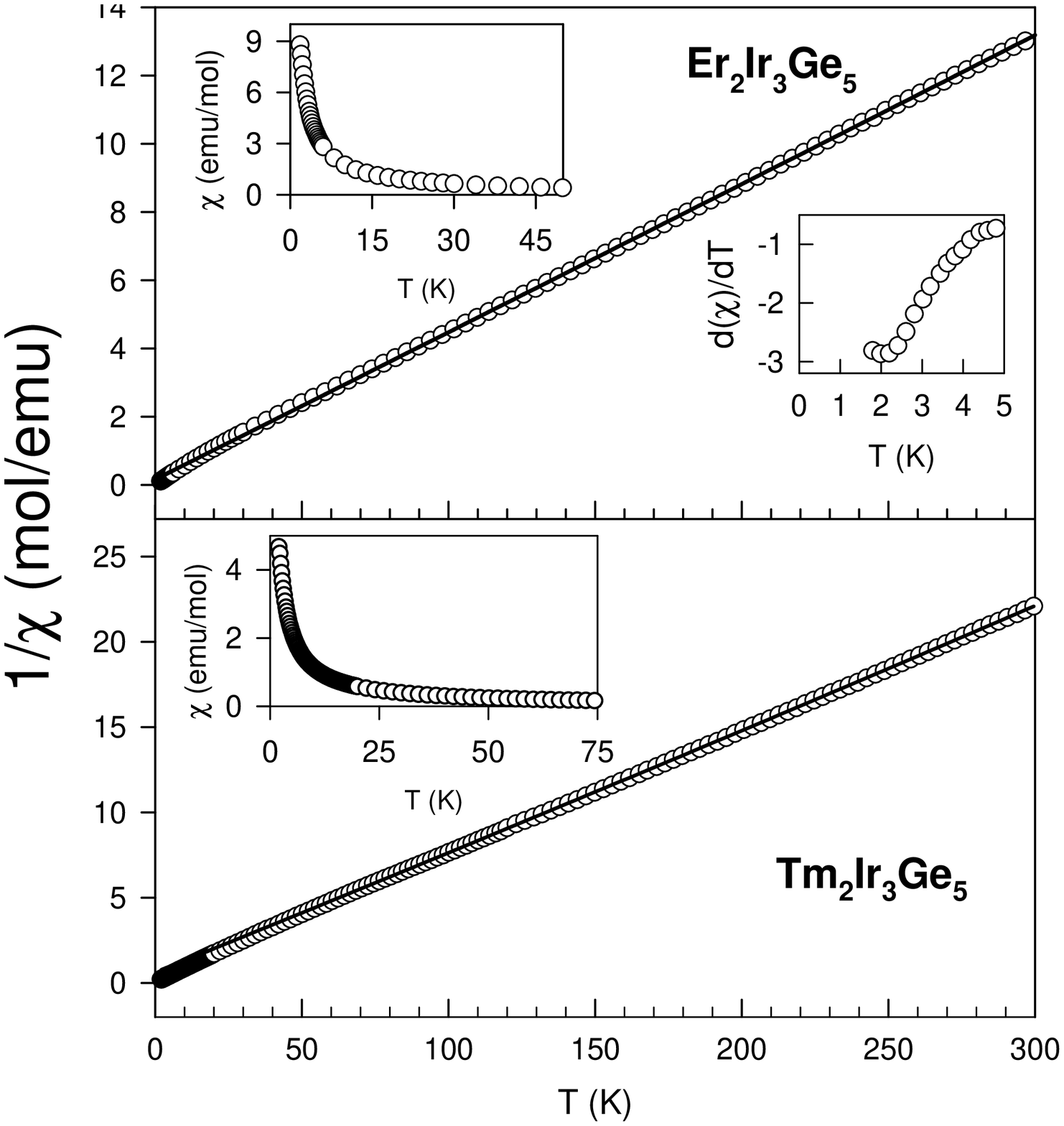}}\end{center}
\caption{Variation of inverse dc susceptibility (1/$\chi$) of 
RE$_2$Ir$_3$Ge$_5$ (RE$=$Er and Tm) from 1.8 to 300~K in a field of 
1~kOe. The inset shows the $\chi$ behavior at low temperatures.
The solid line is a fit to the Curie-Weiss relation 
(see text for details). \label{fsus5}}
\end{figure} 
\begin{figure} 
\begin{center}\resizebox{4.5in}{!}{\includegraphics{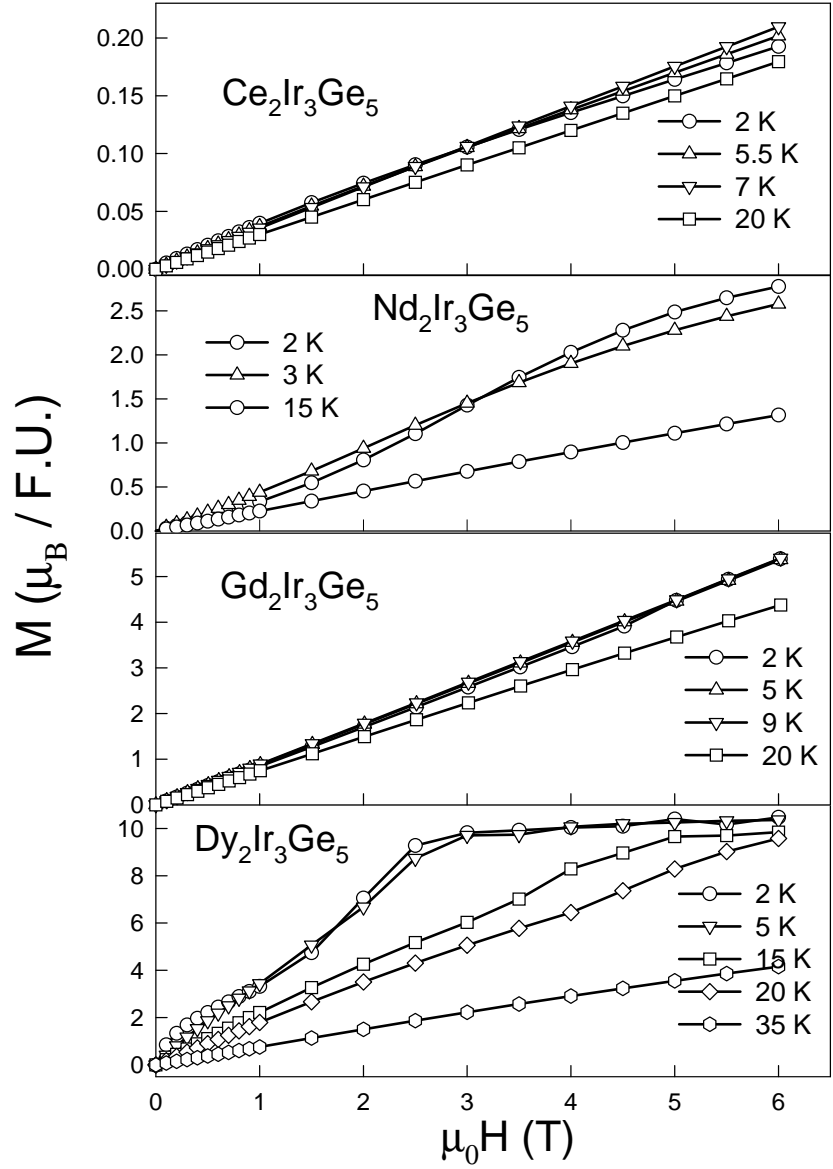}}\end{center}
\caption{Isothermal magnetization (M) of RE$_2$Ir$_3$Ge$_5$ (RE$=$Ce, Nd, Gd
and Dy) vs magnetic field (H) at various  temperatures. The non-linearity in 
M vs H at the lowest temperature agrees with the notion of antiferromagnetic 
ordering of RE$^{3+}$ spins whereas the linear dependence of M on H above T$_N$ (except for Dy$_2$Ir$_3$Ge$_5$) signifies that the sample is in the paramagnetic state at this temperature.
\label{fsus6}}
\end{figure} 
\begin{figure} 
\begin{center}\resizebox{4.5in}{!}{\includegraphics{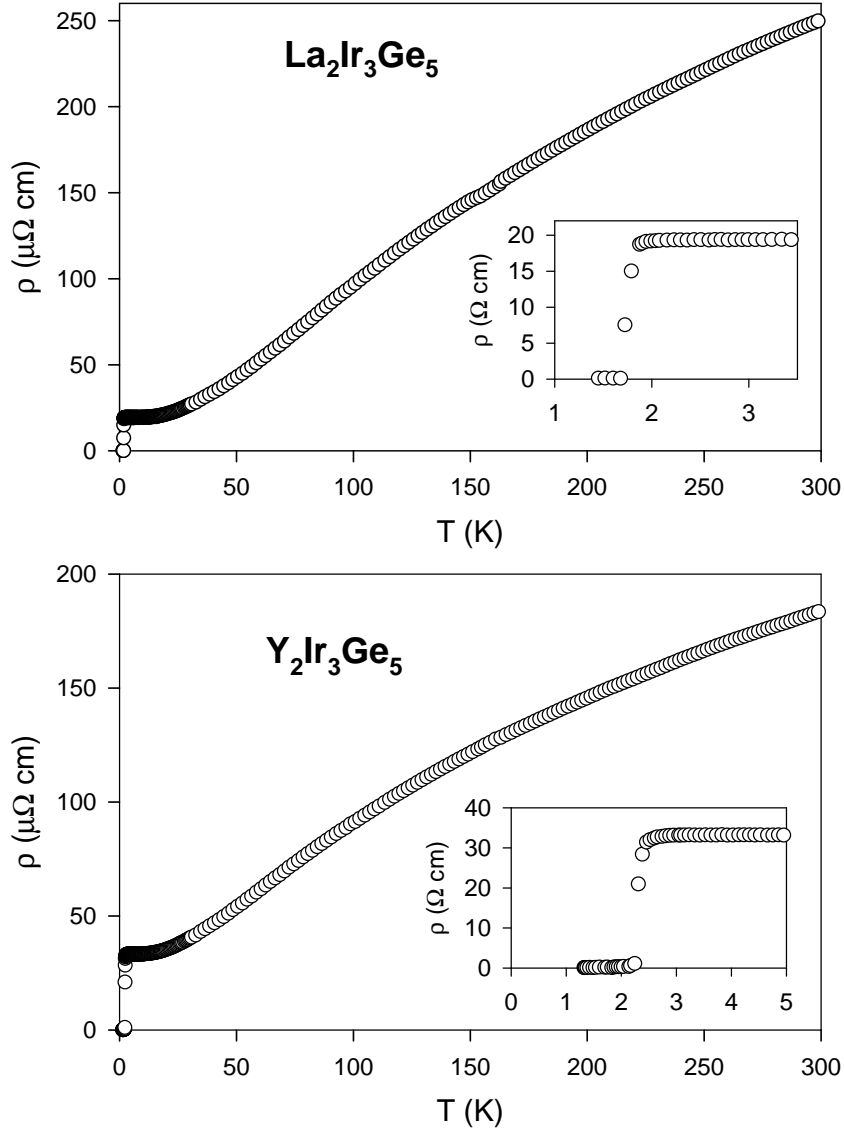}}\end{center}
\caption{Temperature dependence  of resistivity ($\rho$) of 
RE$_2$Ir$_3$Ge$_5$ (RE$=$Y and La) from 1.5 to 300~K. The insets 
show the low temperature $\rho$ data. The superconducting transitions can be clearly seen for both La$_2$Ir$_3$Ge$_5$ and Y$_2$Ir$_3$Ge$_5$. \label{fres1}}
\end{figure}
\begin{figure} 
\begin{center}\resizebox{4.5in}{!}{\includegraphics{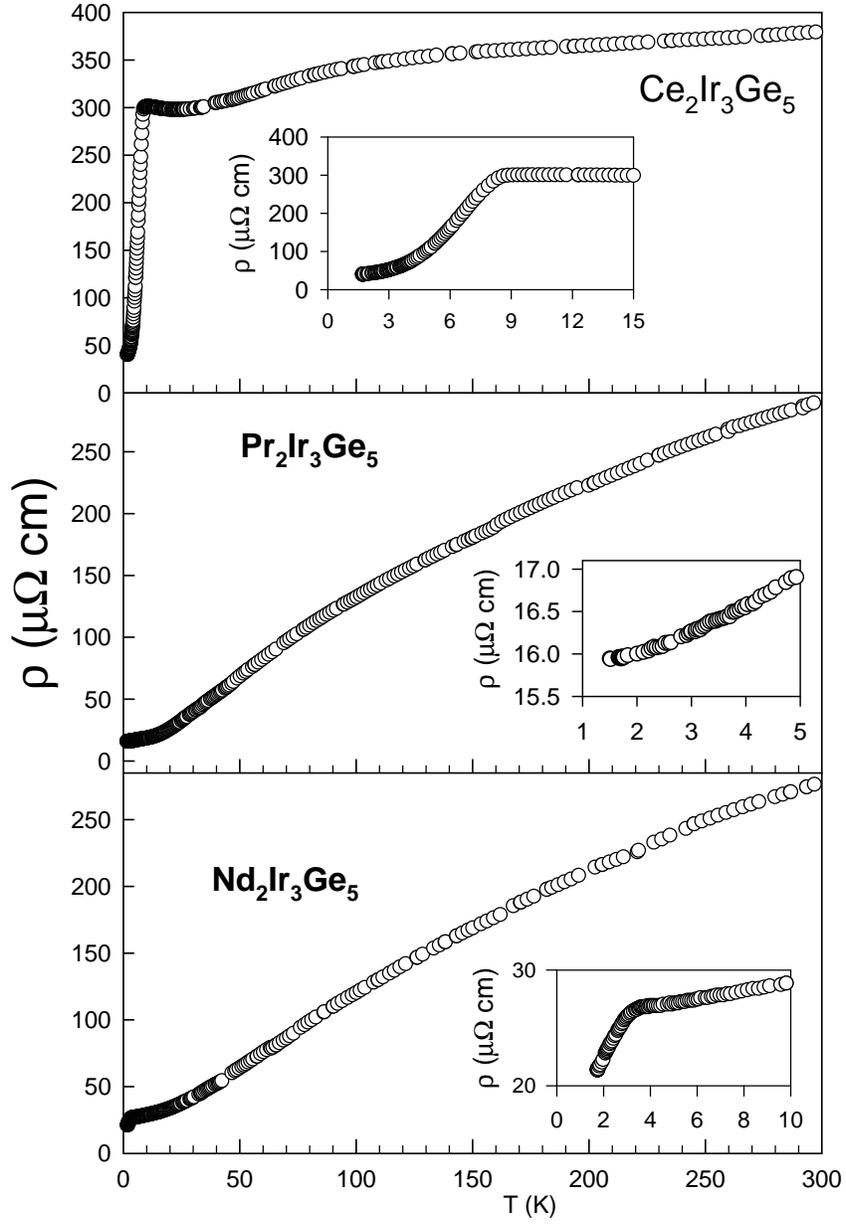}}\end{center}
\caption{Temperature dependence of resistivity ($\rho$) of 
RE$_2$Ir$_3$Ge$_5$ (RE$=$Ce, Pr and Nd) from 1.5 to 300~K. The insets 
show the low temperature $\rho$ vs T data. \label{fres2}}
\end{figure} 
\begin{figure} 
\begin{center}\resizebox{4.5in}{!}{\includegraphics{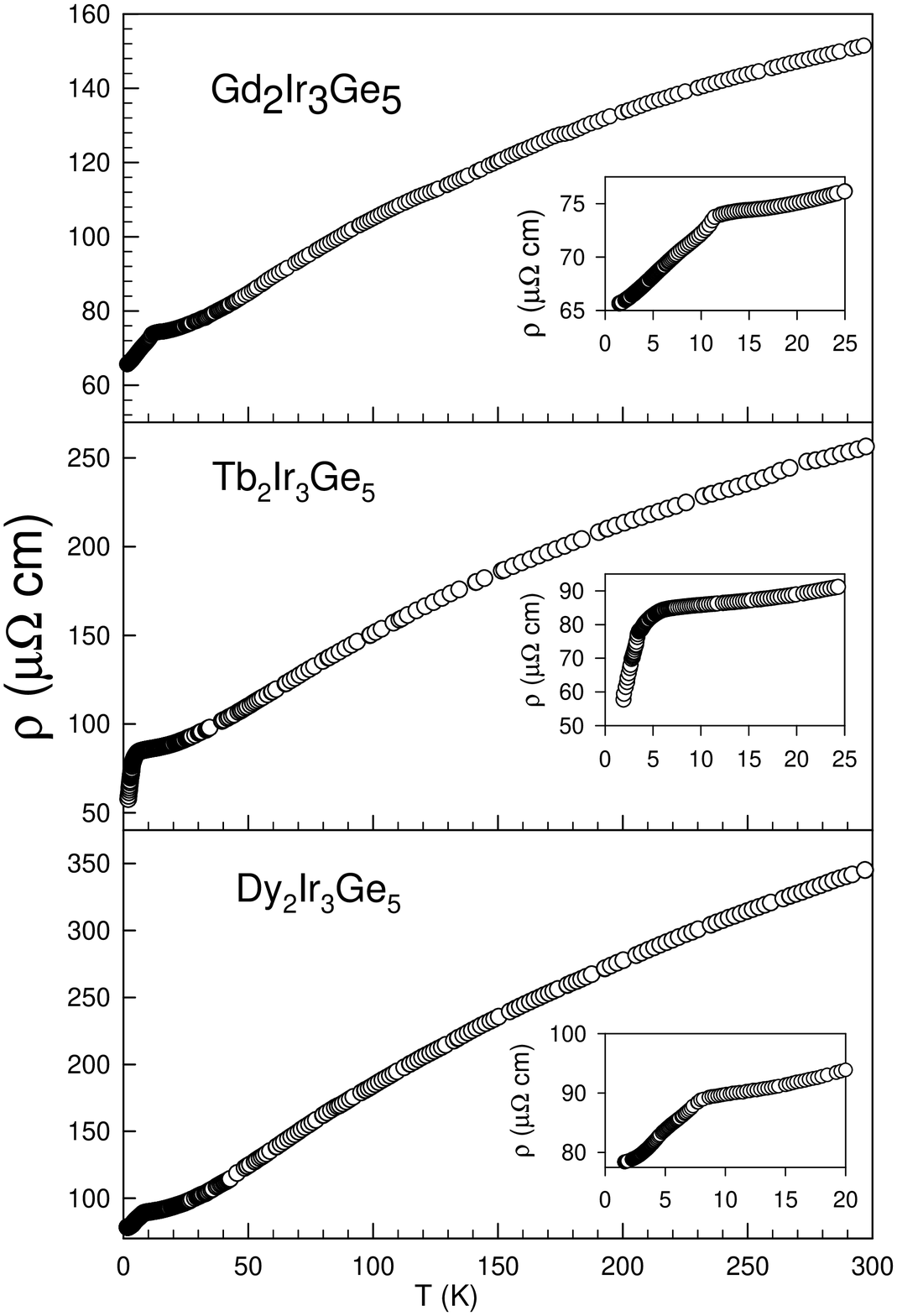}}\end{center}
\caption{Temperature dependence of resistivity ($\rho$) of 
RE$_2$Ir$_3$Ge$_5$ (RE$=$Gd, Tb and Dy) from 1.5 to 300~K. The insets 
show the low temperature $\rho$ vs T data. \label{fres3}}
\end{figure}
\begin{figure} 
\begin{center}\resizebox{4.5in}{!}{\includegraphics{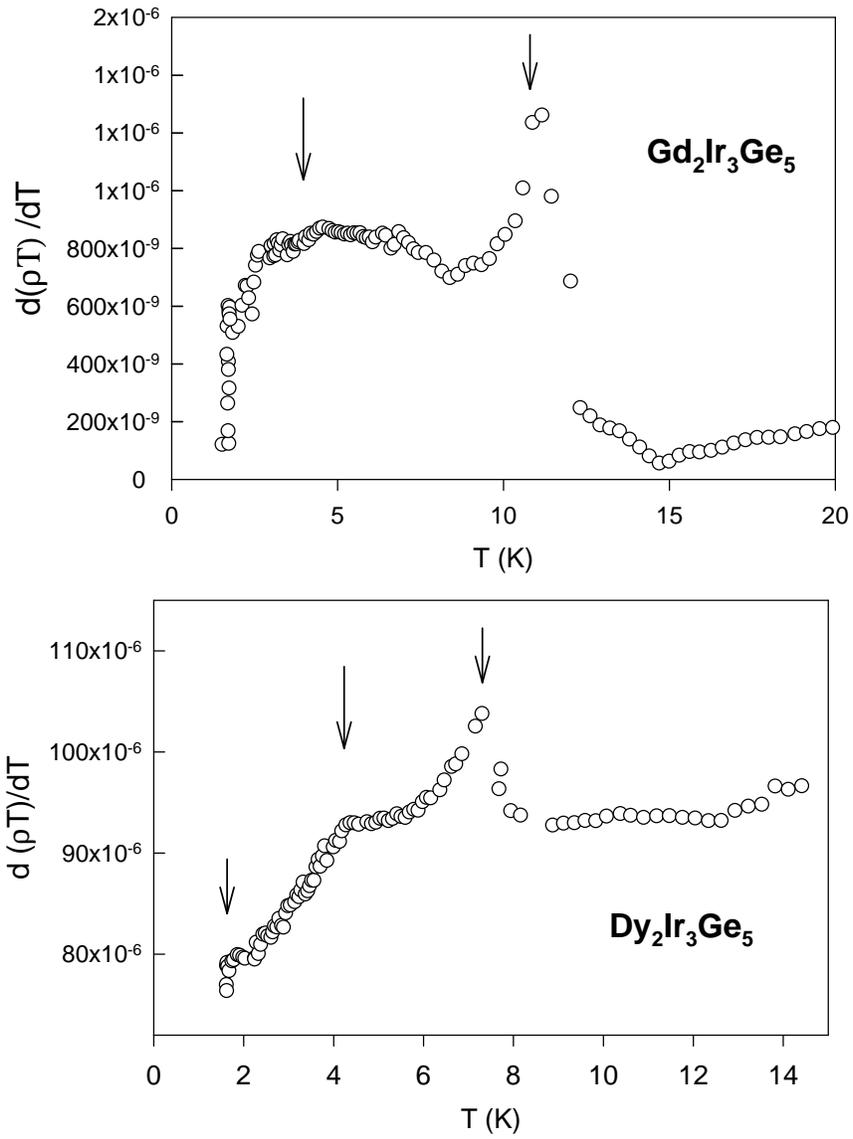}}\end{center}
\caption{Temperature dependence of d$\rho$/dT for Gd$_2$Ir$_3$Ge$_5$ and Dy$_2$Ir$_3$Ge$_5$ . The arrows mark the multiple anomalies for both compounds. \label{fres4}}
\end{figure} 
\begin{figure} 
\begin{center}\resizebox{4.5in}{!}{\includegraphics{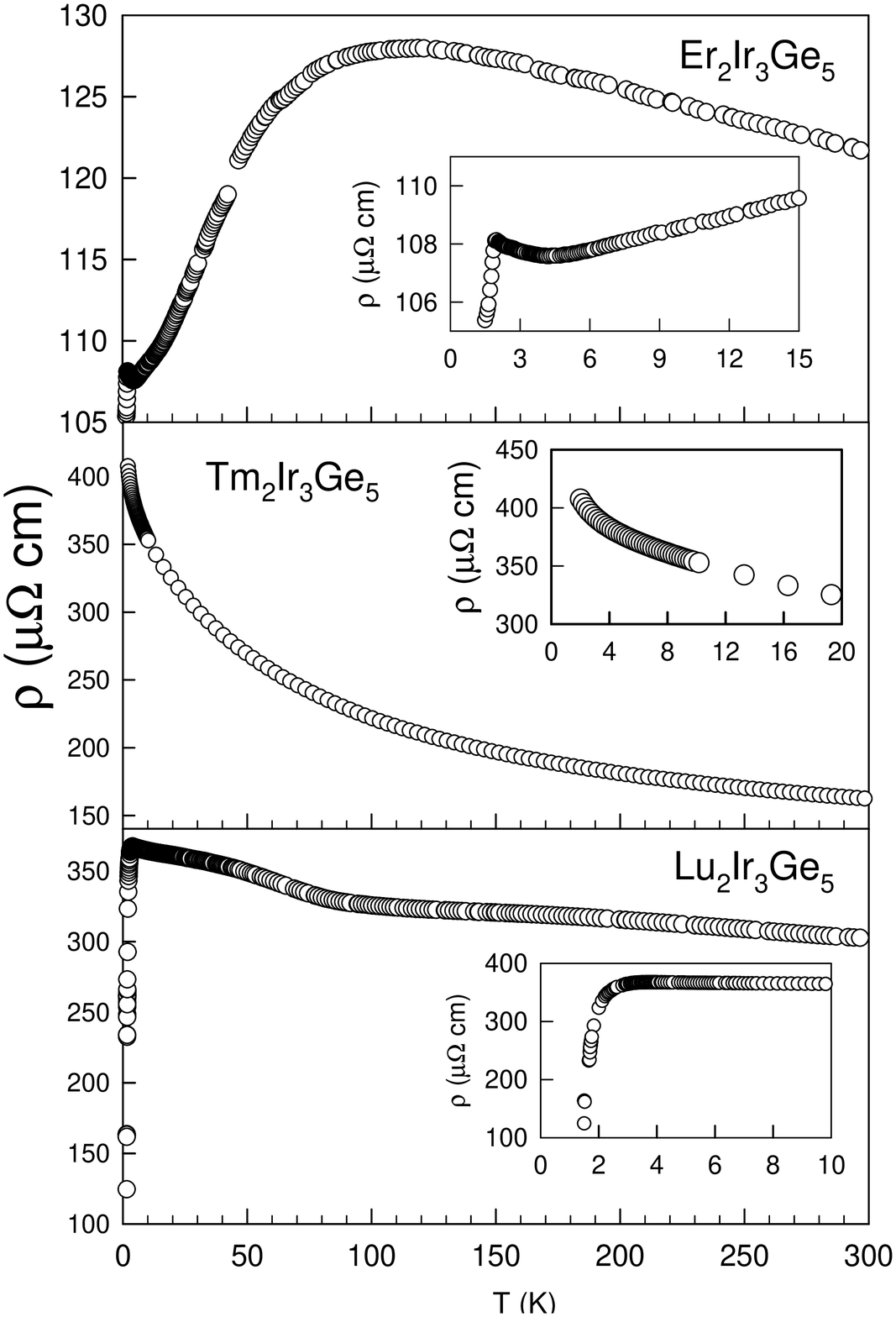}}\end{center}
\caption{Temperature dependence of resistivity ($\rho$) of 
RE$_2$Ir$_3$Ge$_5$ (RE$=$Er, Tm and Lu) from 1.5 to 300~K. The insets 
show the low temperature $\rho$ vs T data. \label{fres5}}
\end{figure} 
\begin{figure} 
\begin{center}\resizebox{4.5in}{!}{\includegraphics{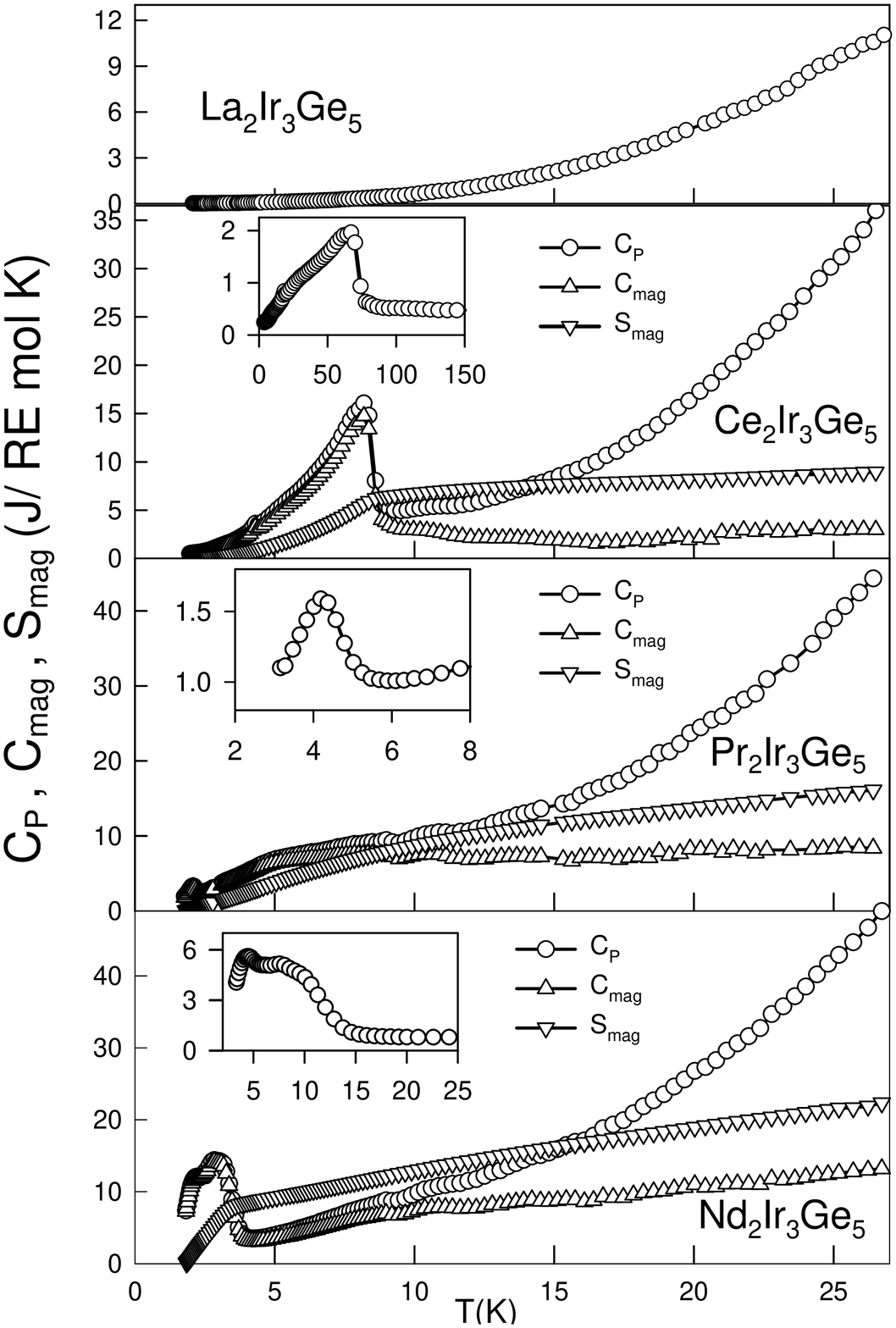}}\end{center}
\caption{Temperature dependence of the heat-capacity (C$_p$) of 
RE$_2$Ir$_3$Ge$_5$ (RE$=$La, Ce, Pr and Nd) from 1.8 to 30~K. The magnetic entropy C$_{mag}$ and the calculated values of the entropy S$_{mag}$ (after the subtraction of the lattice contribution from C$_p$) are also given in the same figure. The inset 
shows the low temperature C$_p$/T vs T$^2$ data.\label{fspecific1}} 
\end{figure}
\begin{figure} 
\begin{center}\resizebox{4.5in}{!}{\includegraphics{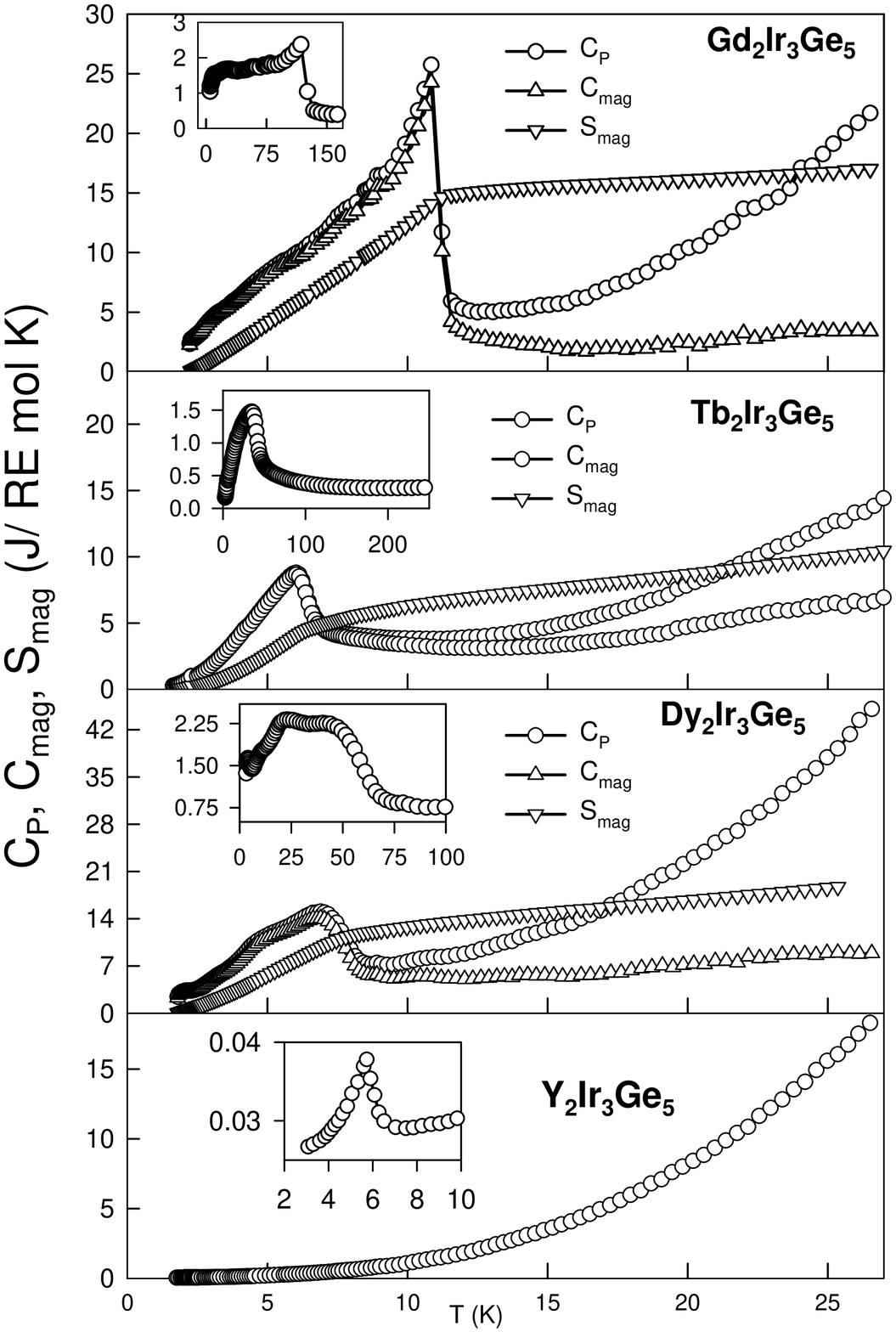}}\end{center}
\caption{Temperature dependence of the heat-capacity (C$_p$) of 
RE$_2$Ir$_3$Ge$_5$ (RE$=$Gd, Tb, Dy and Y) from 1.8 to 30~K. The magnetic entropy C$_{mag}$ and the calculated values of the entropy S$_{mag}$ (after the subtraction of the lattice contribution from C$_p$) are also given in the same figure. The inset 
shows the low temperature C$_p$/T vs T$^2$ data.\label{fspecific2}}  
\end{figure}
\begin{figure} 
\begin{center}\resizebox{4.5in}{!}{\includegraphics{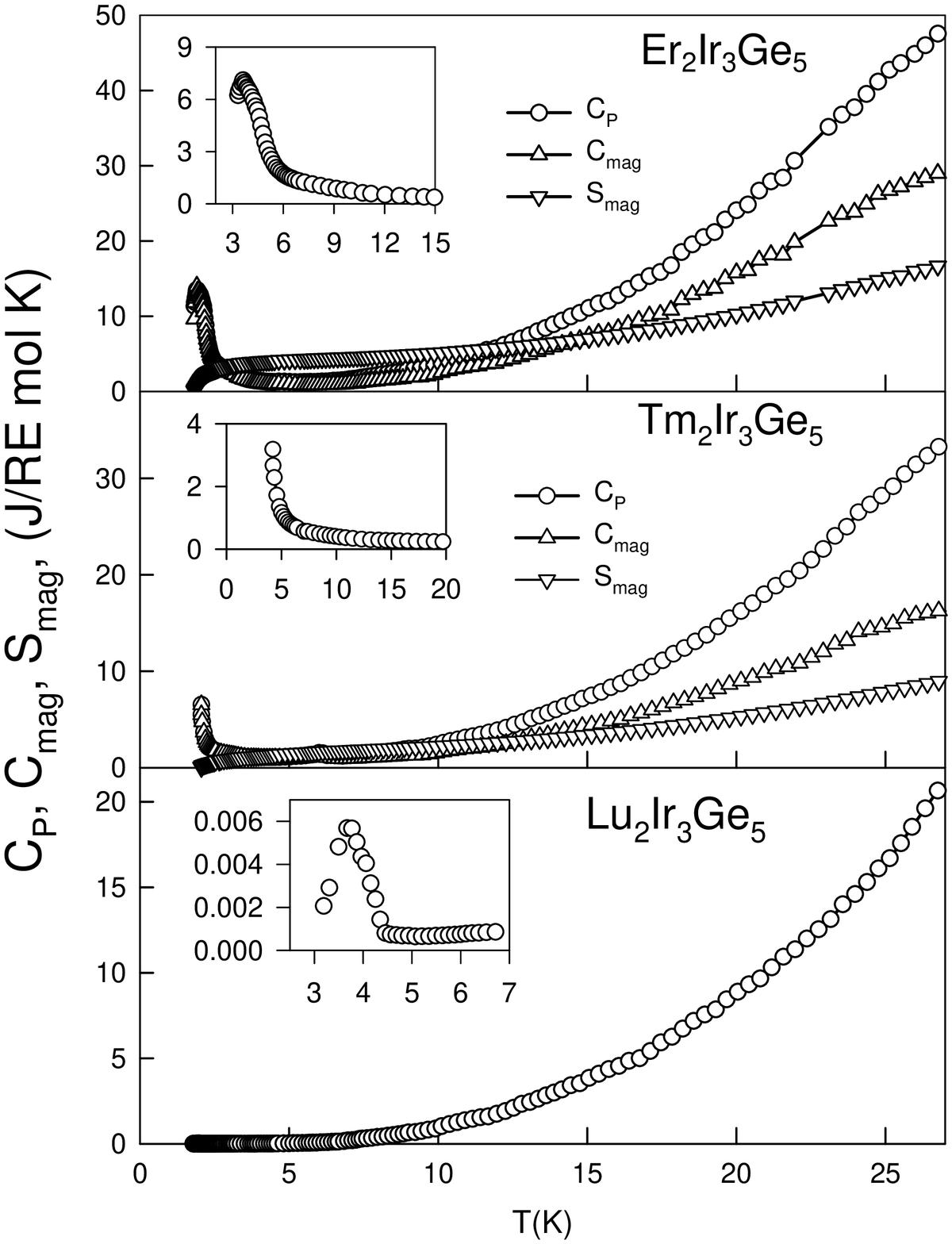}}\end{center}
\caption{Temperature dependence of the heat-capacity (C$_p$) of 
RE$_2$Ir$_3$Ge$_5$ (RE$=$Er, Tm and Lu) from 1.8 to 30~K. The magnetic entropy C$_{mag}$ and the calculated values of the entropy S$_{mag}$ (after the subtraction of the lattice contribution from C$_p$) are also given in the same figure. The inset 
shows the low temperature C$_p$/T vs T$^2$ data.\label{fspecific3}}
\end{figure}
\begin{figure} 
\begin{center}\resizebox{4.5in}{!}{\includegraphics{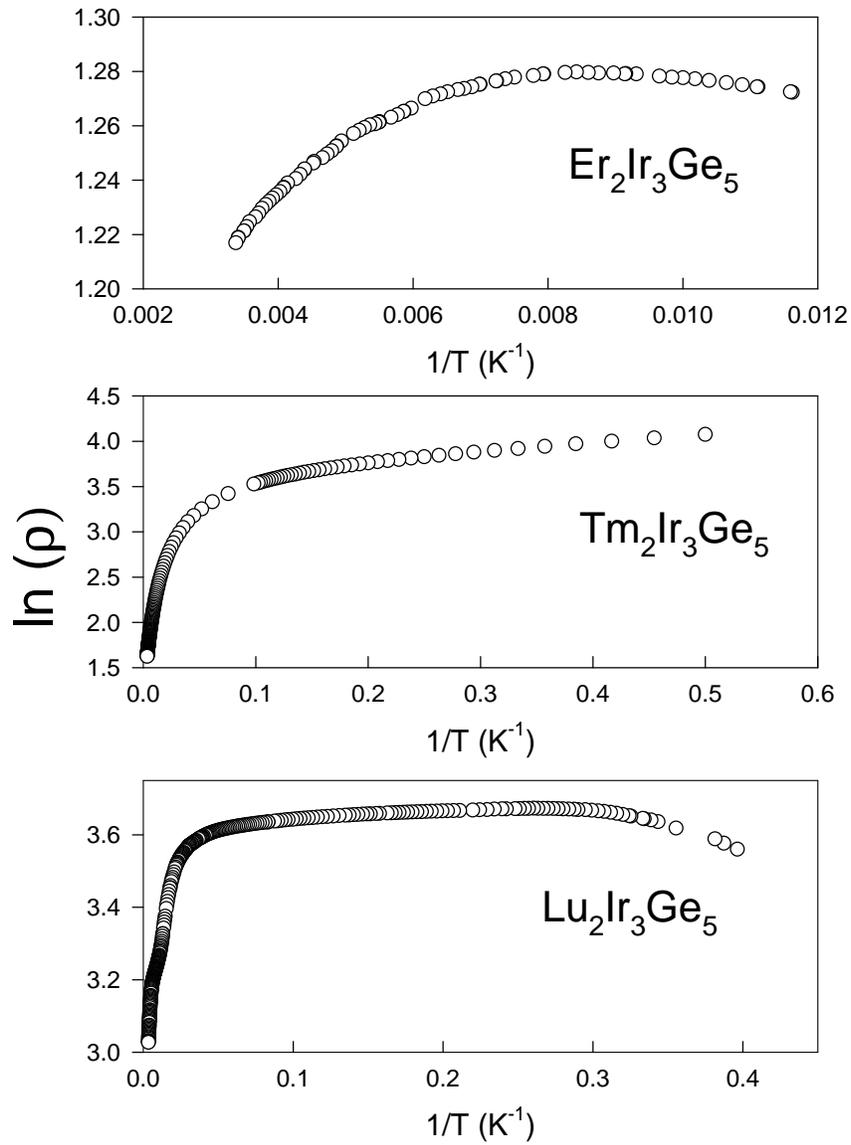}}\end{center}
\caption{Plots of ln($\rho$) vs 1/T for Er$_2$Ir$_3$Ge$_5$, Tm$_2$Ir$_3$Ge$_5$ and Lu$_2$Ir$_3$Ge$_5$ in the temperature range where they show a negetive temperature coefficient.\label{activated}}
\end{figure}
\begin{figure} 
\begin{center}\resizebox{4.5in}{!}{\includegraphics{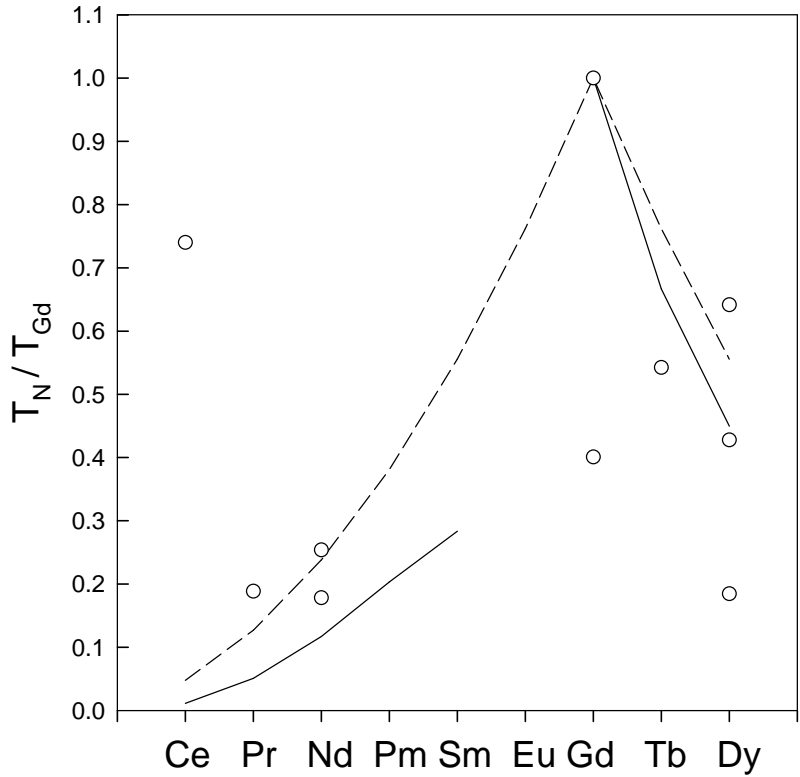}}\end{center}
\caption{Plot of the ordering temperatures of the compounds of the series 
RE$_2$Ir$_3$Ge$_5$ (RE$=$ Ce, Pr, Nd, Tb, Gd, and Dy) normalized to the T$_N$ value for Gd. The dashed lines represents scaling law where only spin quantum number S is used 
whereas the solid lines are for scaling law using total quantum number J 
(de Gennes scaling, see text for details).\label{deGennes}}
\end{figure}
\newpage
\begin{table}
\caption{Lattice parameters of RE$_2$Ir$_3$Ge$_5$}
\begin{tabular}{cccc}
sample&a($\AA$)&b($\AA$)&c($\AA$)\\
\tableline
Y $_2$Ir$_3$Ge$_5$&10.121$\pm$.005&11.802$\pm$.005&5.950$\pm$.005\\
La$_2$Ir$_3$Ge$_5$&10.246$\pm$.005&12.851$\pm$.005&6.104$\pm$.005\\
Ce$_2$Ir$_3$Ge$_5$&10.241$\pm$.005&12.019$\pm$.005&6.094$\pm$.005\\
Pr$_2$Ir$_3$Ge$_5$&10.236$\pm$.005&11.986$\pm$.005&6.074$\pm$.005\\
Nd$_2$Ir$_3$Ge$_5$&10.229$\pm$.005&11.968$\pm$.005&6.064$\pm$.005\\
Gd$_2$Ir$_3$Ge$_5$&10.224$\pm$.005&11.843$\pm$.005&6.032$\pm$.005\\
Tb$_2$Ir$_3$Ge$_5$&10.219$\pm$.005&11.802$\pm$.005&5.997$\pm$.005\\
Dy$_2$Ir$_3$Ge$_5$&10.214$\pm$.005&11.783$\pm$.005&5.984$\pm$.005\\
Er$_2$Ir$_3$Ge$_5$$^*$&19.062$\pm$.005&15.679$\pm$.005&4.633$\pm$.005\\
Tm$_2$Ir$_3$Ge$_5$$^*$&18.998$\pm$.005&15.580$\pm$.005&4.612$\pm$.005\\
Lu$_2$Ir$_3$Ge$_5$$^*$&18.901$\pm$.005&15.382$\pm$.005&4.567$\pm$.005\\

\end{tabular}
~~~\\
$^*$ Different structure (see text).
\label{table 1}
\end{table}
\begin{table}
\caption{Parameters obtained from the high temperature
susceptibility  fit to the Curie-Weiss expression given
by the eqn. (1). $\mu_{th}$ is theoretical free ion value.}
\begin{tabular}{ccccc}
sample & $\chi_0$& $\mu_{eff}$& $\mu_{th}$& $\theta_p$\\
~& $emu/mol~K$& $\mu_B$& $\mu_B$& K\\ \tableline

La$_2$Ir$_3$Ge$_5$&-1.76$\times$10$^{-3}$&-&-&-\\
Ce$_2$Ir$_3$Ge$_5$&3.579$\times$10$^{-4}$&2.535&2.54&-113.2\\
Pr$_2$Ir$_3$Ge$_5$&-2.119$\times$10$^{-4}$&3.567&3.58&-20.69\\
Nd$_2$Ir$_3$Ge$_5$&3.842$\times$10$^{-4}$&3.46&3.62&-7.57\\
Gd$_2$Ir$_3$Ge$_5$&6.498$\times$10$^{-4}$&7.856&7.94&-13.91\\
Tb$_2$Ir$_3$Ge$_5$&-1.375$\times$10$^{-3}$&9.79&9.7&-14.02\\
Dy$_2$Ir$_3$Ge$_5$&-2.332$\times$10$^{-3}$&11.09&10.65&-18.85\\
Er$_2$Ir$_3$Ge$_5$&-6.964$\times$10$^{-4}$&9.64&9.59&-3.813\\
Tm$_2$Ir$_3$Ge$_5$&-1.532$\times$10$^{-3}$&7.6&7.56&-8.99\\
Lu$_2$Ir$_3$Ge$_5$&-2.$\times$10$^{-4}$&-&-&-\\
Y$_2$Ir$_3$Ge$_5$&2.9$\times$10$^{-4}$&-&-&-\\
\end{tabular}
~~~\\
\label{table 2}
\end{table}
\begin{table}
\caption{Transition temperatures T$_p$ (T$_N$ or/and T$_c$) obtained  
from different measurement techniques. Most of them are T$_N$ values 
except for La and Y compounds.}
\begin{tabular}{cccc}
&from $\chi$ &from $\rho$ &from C$_p$\\ \tableline
sample&T$_p$&T$_p$&T$_p$\\ \tableline
&K&K&K\\ \tableline
Y$_2$Ir$_3$Ge$_{5}$&2.6$^*$&2.65$^*$&2.48$^*$\\
La$_2$Ir$_3$Ge$_{5}$&1.74$^*$&1.7$^*$&1.7$^*$\\
Ce$_2$Ir$_3$Ge$_{5}$&8.9&8.5&8.7\\
Pr$_2$Ir$_3$Ge$_{5}$&2.1&-&2.04\\
Nd$_2$Ir$_3$Ge$_{5}$&2.1,2.82&3.&2.08,2.75\\
Gd$_2$Ir$_3$Ge$_{5}$&4.4,11.9&4.2,11.5&4.5,11.21\\
Tb$_2$Ir$_3$Ge$_{5}$&6.4&5.9&6.0\\
Dy$_2$Ir$_3$Ge$_{5}$&2.0,4.3,7.2&1.98,4.7,7.8&2.07,4.79,7.3\\
Er$_2$Ir$_3$Ge$_{5}$&-&$1.97$&1.91\\
Tm$_2$Ir$_3$Ge$_{5}$&-&-&-\\
\end{tabular}
~~~~\\
$^*$  Superconducting transition (T$_c$).
\label{table 3}
\end{table}

\begin{table}
\caption{Parameters obtained from the low temperature resistivity fit of RE$_2$Ir$_3$Ge$_5$}
\begin{tabular}{ccccc}
sample& $\rho_{0}$&a&$n$\\ 
& $\mu\Omega~cm$&$n\Omega~cm/K^n$&n\\ \tableline
La$_2$Ir$_3$Ge$_5$&19.35&.037&3.6\\
Y $_2$Ir$_3$Ge$_5$&33.17&.12&3.28\\
Pr$_2$Ir$_3$Ge$_5$&16.71&6.71&2.42\\
Nd$_2$Ir$_3$Ge$_5$&27.5&45&1.71\\
Gd$_2$Ir$_3$Ge$_5$&73.35&4&2\\
Tb$_2$Ir$_3$Ge$_5$&84.85&7.30&2.12\\
Dy$_2$Ir$_3$Ge$_5$&88.47&9.59&2.12\\
\end{tabular}
\label{table 4}
\end{table}
\begin{table}
\caption{Parameters obtained from the  fit of the high temperature (75-300~K) $\rho$(T) data 
to the parallel resistor model in RE$_2$Ir$_2$Ge$_{5}$. $\theta_D$(obs)
is the value estimated from heat-capacity studies.}
\begin{tabular}{cccccc}
sample& $\rho_{max}$&$\rho_0$&C1&$\theta_D$(fit)&$\theta_D$(obs)\\
& $\mu\Omega~cm$&$\mu\Omega~cm$&$\mu\Omega~cm$&K&K\\ \tableline
Y$_2$Ir$_3$Ge$_5$&335&53&824&376&318\\
La$_2$Ir$_3$Ge$_5$&572&40&1337&433&319\\
Pr$_2$Ir$_3$Ge$_5$&744&48&781&266&289\\
Nd$_2$Ir$_3$Ge$_5$&749&51&865&324&287\\
Gd$_2$Ir$_3$Ge$_5$&258&100&394&214&232\\
Tb$_2$Ir$_3$Ge$_5$&548&99&532&206&247\\
Dy$_2$Ir$_3$Ge$_5$&487&128&1169&362&337\\
\end{tabular}
\label{table 5}
\end{table}
\begin{table}
\caption{Parameters obtained from the  heat-capacity measurements on
 RE$_2$Ir$_3$Ge$_{5}$}
\begin{tabular}{cccccc}
sample&T$_N(K)$& S$_{mag}$(T$_N$$^a$)/R&J&ln(2J+1)&$S_{mag}$(30~K)/R\\ \tableline
Ce$_2$Ir$_3$Ge$_5$&8.7&0.678&$5\over2$&1.79&1.08\\
Pr$_2$Ir$_3$Ge$_5$&2.04&0.05&4&2.19&1.94\\
Nd$_2$Ir$_3$Ge$_5$&2.08,2.75$^*$&1.011&$9\over2$&2.30&2.58\\
Gd$_2$Ir$_3$Ge$_5$&4.5,11.21$^*$&1.76&$7\over2$&2.08&2.02\\
Tb$_2$Ir$_3$Ge$_5$&6.0&0.59&6&2.57&1.53\\
Dy$_2$Ir$_3$Ge$_5$&2.07,4.79,7.3 $^*$&1.4&$15\over2$&2.77&2.25\\
Er$_2$Ir$_3$Ge$_5$&1.91&0.45&$15\over2$&2.773&1.82\\
Tm$_2$Ir$_3$Ge$_5$&2.3&0.144&6&2.56&1.10\\
\end{tabular}
~~~\\
$^*$ Multiple transitions.\\
$^a$ Transition temperatures are determined from C$_p$ data.
\label{table 6}
\end{table}
\end{document}